\documentclass[12pt,nohyper,letterpaper]{JHEP3}         
\usepackage{amssymb,amsfonts}

\usepackage{epsfig}

\def\be{\begin{eqnarray}}
\def\ee{\end{eqnarray}}
\newcommand{\nn}{\nonumber}
\newcommand\para{\paragraph{}}
\newcommand{\ft}[2]{{\textstyle\frac{#1}{#2}}}
\newcommand{\eqn}[1]{(\ref{#1})}

\def\Dslash{\,\,{\raise.15ex\hbox{/}\mkern-12mu D}}
\def\Dbarslash{\,\,{\raise.15ex\hbox{/}\mkern-12mu {\bar D}}}
\def\delslash{\,\,{\raise.15ex\hbox{/}\mkern-9mu \partial}}
\def\delbarslash{\,\,{\raise.15ex\hbox{/}\mkern-9mu {\bar\partial}}}
\def\pslash{\,\,{\raise.15ex\hbox{/}\mkern-9mu p}}
\def\calDslash{\,\,{\raise.15ex\hbox{/}\mkern-12mu {\cal D}}}

\newcommand{\D}{{\cal D}}

\newcommand{\Tr}{{\rm Tr}}

\newcommand{\hW}{\hat{\cal W}}

\def\lae{\mathrel{\mathop{\smash{\lower .5 ex \hbox{$\stackrel<\sim$}}}}}
\def\lae{\mathrel{\mathop{\smash{\lower .5 ex \hbox{$\stackrel>\sim$}}}}}

\title{Heterotic Vortex Strings}
\author{Mohammad Edalati\\
Department of Physics, \\
University of Cincinnati, \\
P. O. Box 210011,  \\
Cincinnati, OH 45221-0011, USA \\
{\tt edalati@physics.uc.edu}}

\author{David Tong\\
Department of Applied Mathematics and Theoretical Physics, \\
University of Cambridge, \\
Cambridge, CB3 0WA, UK\\{\tt d.tong@damtp.cam.ac.uk}}

\abstract{We determine the low-energy ${\cal N}=(0,2)$ worldsheet
dynamics of vortex strings in a large class of non-Abelian ${\cal
N}=1$ supersymmetric gauge theories.}

\begin{document}
\pagestyle{plain} \setcounter{page}{1}
\newcounter{bean}
\baselineskip16pt \setcounter{section}{0}

\tableofcontents
\section{Introduction}

Vortex strings provide a map between four-dimensional non-Abelian
gauge theories and two-dimensional sigma-models. The
four-dimensional theories in question have a  $U(N_c)$ gauge group
and a sufficient number of scalar fields to allow complete gauge
symmetry breaking, so that the system lies in the Higgs phase.
Theories with this property admit vortex strings. The embedding of
the vortex within the non-Abelian gauge group endows the string
with a number of orientation modes which parameterize the complex
projective space ${\bf CP}^{N_c-1}$. Further bosonic and fermionic
zero modes of the vortex live in line bundles over ${\bf
CP}^{N_c-1}$. In this manner, the low-energy dynamics of a single,
straight, infinite vortex string is described by some variant of
the ${\bf CP}^{N_c-1}$ sigma-model living on the $d=1+1$
dimensional worldsheet \cite{vib,auzzi}.

\para
When the four-dimensional gauge theory has ${\cal N}=2$
supersymmetry, a pleasing story emerges. The strings are
$1/2$-BPS, ensuring that the worldsheet dynamics inherits ${\cal
N}=(2,2)$ supersymmetry. It was shown in \cite{sy,vstring},
following earlier work of \cite{nick,dht}, that the quantum
dynamics of the worldsheet theory encodes quantitative information
about the quantum dynamics of the parent four-dimensional theory,
including the Seiberg-Witten curve and the exact BPS  mass
spectrum. More recently, the correspondence was extended to
superconformal points, with a matching between the scaling
dimensions of chiral primary operators in the four-dimensional
bulk and on the worldsheet \cite{scvs}. For a review of the
classical and quantum dynamics of these strings, see \cite{tasi}.

\para
The purpose of this paper is to present a detailed study of the
classical dynamics of vortex strings in ${\cal N}=1$
four-dimensional gauge theories. For certain choices of parameters
the strings once again preserve $1/2$ of supersymmetry, now
guaranteeing ${\cal N}=(0,2)$ supersymmetry on the worldsheet. For
this reason, we refer to vortices in ${\cal N}=1$ theories as
``heterotic vortex strings". We will determine the explicit ${\cal
N}=(0,2)$ ${\bf CP}^{N_c-1}$ sigma-models, and their variations,
which describe the low-energy dynamics of vortex strings in a
large class of ${\cal N}=1$ gauge theories\footnote{Vortex strings
in various non-Abelian theories with less supersymmetry were
previously studied in [9-13]
and in some cases qualitative agreement was found between the
dynamics of the worldsheet theory and the bulk. We will comment
more on the relationship of our work to some of these papers in
Section 4.}.

\para
The paper is organized as follows: Section 2 contains a detailed
discussion of the ${\cal N}=(2,2)$ worldsheet dynamics of vortex
strings in ${\cal N}=2$ four-dimensional theories. This section is
mostly a review of previous work, although explicit expressions
for bosonic and fermionic zero modes are provided which generalize
results in the literature from $U(2)$ gauge theories to $U(N_c)$
gauge theories. Particular attention is paid to the chirality of
different fermionic zero modes since this will prove important in
later sections. Section 3 also contains review material,
describing the basics of the superfield formalism for ${\cal
N}=(0,2)$ supersymmetry in $d=1+1$ dimensions.

\para
The meat of the paper is in Section 4. We consider two different
classes of  deformations, each of  which breaks the
four-dimensional supersymmetry from ${\cal N}=2$ to ${\cal N}=1$
through  the introduction of a superpotential for the adjoint
chiral multiplet. In each case, we show that there is a unique
${\cal N}=(0,2)$ worldsheet theory which correctly captures all
BPS properties of the vortex string and predicts the interaction
of fermionic zero modes. We also include an appendix which
collates the notation for bulk and worldsheet fields used
throughout the paper.

\section{The ${\cal N}=(2,2)$ Dynamics of Vortex Strings}

In this section we review the dynamics of vortex strings in
four-dimensional gauge theories with ${\cal N}=2$ supersymmetry.
The vortices are $1/2$-BPS, ensuring that the $d=1+1$ dimensional
worldsheet dynamics of the string inherits ${\cal N}=(2,2)$
supersymmetry.

\subsection{The Four-Dimensional Theory}

Our starting point is the  $d=3+1$, ${\cal N}=2$ supersymmetric
$U(N_c)$ gauge theory, with $N_f$ flavors transforming in the
fundamental representation\footnote{Conventions: We pick Hermitian
generators $T^m$ with Killing form $\Tr\, T^mT^n= \ft12
\delta^{mn}$. We write the gauge field as $A_\mu=A_\mu^mT^m$ and
$F_{\mu\nu}=\partial_\mu A_\nu-\partial_\nu A_\mu -
i[A_\mu,A_\nu]$. Fundamental  covariant derivatives are ${\cal
D}_\mu Q=\partial_\mu Q-iA_\mu Q$; adjoint covariant derivatives
are ${\cal D}_\mu A =
\partial_\mu A - i [A_\mu,A]$. Our summation
conventions are inconsistent: a sum over repeated indices is
usually left implicit unless there is some ambiguity or a point
that requires emphasis.}. We describe the theory in the language
of four-dimensional ${\cal N}=1$ superfields. The ${\cal N}=2$
vector multiplet consists of an ${\cal N}=1$ vector multiplet $V$
and an ${\cal N}=1$ adjoint chiral multiplet $A$. Similarly, each
flavor hypermultiplet splits into two chiral multiplets, $Q_i$ and
$\tilde{Q}_i$ where $i=1,\ldots, N_f$ is the flavor index. Each
$Q_i$ transforms in the fundamental ${\bf N}_c$ of the gauge
group, while each $\tilde{Q}_i$ transforms in the anti-fundamental
$\bar{\bf N}_c$. We denote the complexified gauge coupling of the
theory as
\be \tau=\frac{2\pi i}{e^2} + \frac{\theta}{2\pi}\ .
\label{tau}\ee
The four
dimensional theory has the usual superpotential required for
${\cal N}=2$ supersymmetry,
\be{\cal W}_{{\cal
N}=2}=\sqrt{2}\,\sum_{i=1}^{N_f}\tilde{Q}_iAQ_i\ .\label{4dn2}\ee
The scalar potential of the theory is dictated by the D-term and
the F-terms arising from this  superpotential. In components it is
given by,
\be V_{4d}&=& \frac{e^2}{2}\Tr(\,\sum_{i=1}^{N_f}\,Q_iQ_i^\dagger
- \tilde{Q}_i\tilde{Q}_i^\dagger - v^2\,1_{N_c})^2 +e^2\Tr|\,
\sum_{i=1}^{N_f}\tilde{Q}_iQ_i|^2\nn\\
&&+\sum_{i=1}^{N_f}\left(Q_i^\dagger \{A,A^\dagger\}Q_i  +
\tilde{Q}_i\{A,A^\dagger\}\tilde{Q}_i^\dagger\right) +
\frac{1}{2e^2}\Tr|[A,A^\dagger]|^2\label{v4dn2}\ee
where we have taken the liberty of denoting the component scalar
fields by the same Roman letter as the superfield in which they
reside. We have included a D-term Fayet-Iliopoulos (FI) parameter
$v^2$ for the central $U(1)\subset U(N_c)$. This is consistent
with ${\cal N}=2$ supersymmetry and forces the theory into the
Higgs phase, with $Q_i$ gaining a vacuum expectation value (vev).
For $N_f<N_c$, the rank condition ensures the D-term cannot vanish
and there is no supersymmetric ground state. We do not consider
this case. When $N_f>N_c$, the D-term and F-term conditions do not
fix the vevs of $Q_i$ and $\tilde{Q}_i$ completely and there is a
Higgs branch of vacua; we shall discuss this situation in Section
\ref{add}. For now we restrict attention to the case $N_f=N_c$ for
which there is a unique supersymmetric ground state in which the
gauge group is completely broken. Up to a gauge transformation the
ground state is given by
\be Q^a_{\ i}=v\delta^a_{\ i}\ \ \ ,\ \ \
\tilde{Q}_i=A=0\label{vac1}\ee
where $a=1,\ldots, N_c$ is the color index. The theory lies in the
color-flavor locked phase, with the vacuum expectation value
preserved by a simultaneous gauge and flavor rotation. The
symmetry breaking pattern is thus broken to the diagonal
combination of the two (recall that we are looking at the theory with $N_f=N_c$)
\be U(N_c)\times SU(N_f)\rightarrow SU(N_c)_{\rm
diag}\ .\label{breaking}\ee

\subsection{The Vortex}

The central $U(1)\subset U(N_c)$ does not survive the symmetry
breaking \eqn{breaking}, a fact which provides sufficient topology
to ensure the presence of vortex strings in the theory \cite{no}.
These vortices preserve $1/2$ of the supersymmetry, ensuring
${\cal N}=(2,2)$ supersymmetric dynamics on their $d=1+1$
dimensional worldvolume. Infinite, straight strings oriented in
the $x^3$ direction satisfy the first-order equations,
\be F_{12}=e^2(\sum_{i=1}^{N_f}Q_iQ_i^\dagger - v^2\,1_{N_c})\nn\\
{\cal D}_{{z}}Q_i\equiv \frac{1}{2}({\cal D}_1Q_i-i{\cal D}_2Q_i)
=0 \label{vort}\ee
where $z=x^1+ix^2$ parameterizes the transverse plane.
These are the non-Abelian vortex equations.
Solutions to these equations have tension
\be T_k=2\pi k v^2 \ee
where $k =-\Tr\,\int (F_{12}/2\pi) \in {\bf Z}^+$ is the winding
number.

\para
\EPSFIGURE{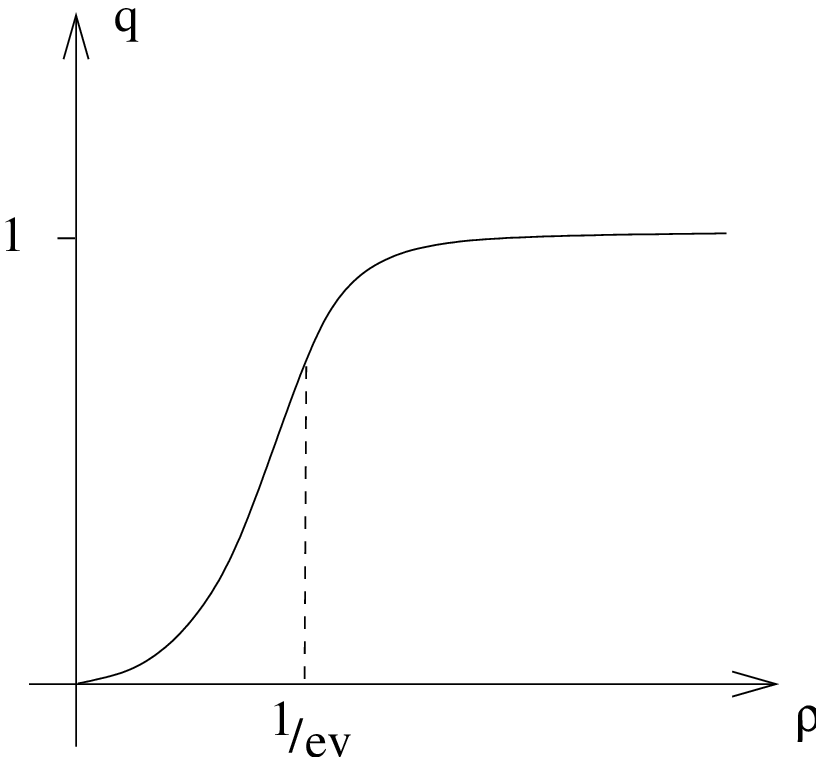,height=110pt}{} Solutions to the vortex
equations with winding number $k$ have $2kN$ bosonic collective
coordinates. For a single $k=1$ vortex, they break down as
follows: there are $2$ collective coordinates corresponding to the
position of the string in the $z=x^1+ix^2$ plane. The remaining
$2(N-1)$ collective coordinates are Goldstone modes arising from
the action of the surviving symmetry \eqn{breaking} on the vortex
string. They parameterize $SU(N_c)/[SU(N_c-1)\times U(1)] \cong
{\bf CP}^{N_c-1}$ \cite{vib,auzzi}.

\para
An explicit realization of the orientational modes is most simply
given in {\it singular gauge} in which $Q$ does not wind
asymptotically, with the flux instead arising from a singular
gauge potential \cite{auzzi}. Suppose that the Abelian $N_c=1$
vortex equations are solved by two profile functions $q(\rho)$ and
$a(\rho)$, where $\rho=\sqrt{(x^1)^2+(x^2)^2}$ is the radial
distance from the string
\be Q_{\rm Abelian} = v q(\rho)\ \ \ \ {\rm and}\ \ \ \ (A_z)_{\rm
Abelian}= -i\bar{z} a(\rho)\ .  \ee
Here the complexified gauge connection is $A_z=\ft12 (A_1-iA_2)$.
Plugging this ansatz into the vortex equations gives two first
order ordinary differential equations,
\be  q'=2\rho aq\ \ \ {\rm  and} \ \ \ 4a+2\rho a'=e^2v^2
(q^2-1)\ee
with prime denoting the derivative with respect to $\rho$. These
equations are known to admit a unique solution satisfying the
appropriate boundary conditions,
\be q(\rho)\rightarrow \left\{\begin{array}{c} 1 \\ 0
\end{array}\right.\ \ \ \ ,\ \ \ \ \ a(\rho)\rightarrow
\left\{\begin{array}{lc} 0 \ \ \ \ \ & {\rm as}\ \rho\rightarrow
\infty
\\ 1/2\rho^2 & {\rm as}\ \rho\rightarrow 0\ .
\end{array}\right.\ee
However, the solution does not have a simple analytic form. A
sketch of the profile $q(\rho)$ is shown in figure 1.

\para
With the $k=1$ Abelian vortex solution in hand, one may simply
construct a solution to the non-Abelian equations by embedding
thus,
\be Q^a_{\ i}= \left(\frac{\phi^a\bar{\phi}_i}{r}\right)\,
v[q(\rho)-1]+v\delta^a_{\ i} \ \ \ \ {\rm and}\ \ \ \ (A_z)^a_{\
b} = -i\bar{z}a(\rho)\,\left(\frac{\phi^a\bar{\phi}_b}{r}\right) \
.\label{solutions}\ee
The $\phi^a \in {\bf C}^{N_c}$ define the orientation of the
vortex in the gauge and flavor groups. In order that this reduce
to the Abelian solution, we require
\be \sum_{a=1}^{N_c} |\phi^a|^2=r \label{phiphi}\ee
with $r$ a constant that will be fixed shortly. The solutions
\eqn{solutions} are invariant under the simultaneous rotation,
\be \phi^a\rightarrow e^{i\alpha}\phi^a\ .\label{iden}\ee
The $\phi^a$, subject to the constraint \eqn{phiphi} and
identification \eqn{iden}, provide homogeneous coordinates on
${\bf CP}^{N_c-1}$. The $SU(N_c)$ symmetry of four-dimensions
descends to the vortex string, with the $\phi^a$ transforming in
the fundamental representation. This ensures that the ${\bf
CP}^{N_c-1}$ is endowed with the symmetric Fubini-Study metric.
The K\"ahler class of this space is $r$.

\para
A comment on notation: since $N_f=N_c$, both $Q^a_{\ i}$ and
$(A_z)^a_{\ b}$ are $N_c\times N_c$ matrices. In what
follows, we shall often neglect to write the indices on both. In
this notation, $Q$ is a matrix on which gauge rotations $U\in
U(N_c)$ act from the left, while flavor rotations $V\in SU(N_f)$
act from the right, so that $Q\rightarrow UQV^\dagger$.

\subsubsection{Bosonic Zero Modes}

For general winding number $k$, the vortex zero modes are defined
to be solutions to the linearized vortex equations,
\be {\cal D}_z\delta A_{\bar{z}}-{\cal D}_{\bar{z}}\delta A_z &=&
\frac{ie^2}{2}(\delta Q Q^\dagger + Q\delta Q^\dagger) \nn\\
{\cal D}_z \delta Q &=& i\delta A_z Q\ .\ee
These are to be supplemented with a suitable gauge fixing
condition which is derived from Gauss' law and reads
\be {\cal D}_z\delta A_{\bar{z}}+{\cal D}_{\bar{z}}\delta A_z =
-\frac{ie^2}{2}(\delta Q Q^\dagger - Q\delta Q^\dagger)\ .\ee
This gauge fixing condition combines with the first of the
linearized vortex equations to leave us with two, complex, first
order equations to be solved around the background of a fixed
vortex configuration,
\be 2{\cal D}_{\bar{z}}\delta A_z &=& -ie^2\delta Q Q^\dagger
\nn\\ {\cal D}_z \delta Q &=& i\delta A_z Q\label{bogzero}.\ee
We now derive the solutions to these equations that arise from the
symmetries of the system.

\subsubsection*{Translational Mode}

For any winding number $k$, the two translational modes are always
given by
\be \delta A_z=F_{\bar{z}z}\ \ \ {\rm and}\ \ \ \delta Q={\cal
D}_{\bar{z}} Q \label{transzero}\ee
which can be checked to satisfy \eqn{bogzero} using the fact that
the background fields obey the second order equations of motion.

\subsubsection*{Orientational Modes}

The zero modes corresponding to orientation are only slightly more
complicated. In general they can be written as
\be \delta A_z &=& {\cal D}_z\Omega \nn\\
\delta Q &=& i(\Omega Q - Q\hat{\Omega}).\label{orientzero}\ee
Here $\Omega(x)$ is an infinitesimal gauge rotation, while
$\hat{\Omega}$ is an infinitesimal flavor rotation. Since only the
diagonal subgroup \eqn{breaking} of these is preserved in the
vacuum, we require that $\Omega(x)\rightarrow \hat{\Omega}$ as
$x\rightarrow \infty$. In terms of our orientation coordinates
$\phi^i$, this diagonal rotation can be written as,
\be \hat{\Omega}^i_{\ j}=-i\left[\delta\phi^i\bar{\phi}_j - \phi^i
\delta \bar{\phi}_j - 2iu \phi^i\bar{\phi}_j\right]\ee
which holds for any $u$. Requiring that $\hat{\Omega} \in su(N_c)$
fixes $u$ to be
\be u = -i\bar{\phi}_i\,\delta\phi^i\ .\ee
Later $u$ will become a gauge field on the worldsheet whose role
is to implement the identification \eqn{iden}. For now, we can
treat $u$ as a connection and introduce the covariant variation
$\nabla\phi^i=\delta\phi^i-iu\phi^i$ which satisfies
$\nabla\phi^i\cdot\bar{\phi}_i=0$. In this notation
\be \hat{\Omega}^i_{\ j} =
-i[(\nabla\phi^i)\bar{\phi}_j-\phi^i\nabla\bar{\phi}_j]. \ee
The zero mode equations \eqn{bogzero} translate to the requirement
that $\Omega(x)$ satisfy the second order differential equation
\be {\cal D}^2\Omega = e^2 \left[\{\Omega,
QQ^\dagger\}-2Q\hat{\Omega} Q^\dagger\right].\label{hard}\ee
Everything above holds for arbitrary winding number $k$. For a
single vortex, with $k=1$, the solution to \eqn{hard} was provided
in \cite{others} (see equation (28) of that paper) and depends
only on the profile function $q(\rho)$ of the
vortex\footnote{Equation \eqn{osol}  solves \eqn{hard} by virtue
of the vortex profile obeying the second order equation
$4\partial_z\partial_{\bar{z}}q-4a^2\rho^2q=e^2v^2q(q^2-1)$.}
\be \Omega(\rho) = q(\rho)\,\hat{\Omega}.\label{osol}\ee
Using the solution \eqn{osol}, we can now be more explicit about
the orientation zero modes for a single vortex. Making use of the
vortex equations \eqn{vort}, we find
\be (\delta A_z)^a_{\ b} &=& -2i(\partial_z q)
\,(\nabla\phi^a)\bar{\phi}_b \nn\\ \delta Q^a_{\ i} &=&
v(q^2-1)\,(\nabla\phi^a)\bar{\phi}_i.\ee

\subsection{Fermions}

We now turn to a study of the fermionic zero modes \cite{jackr}. We start by
describing the Dirac equations in four-dimensions and their
solutions for a single $k=1$ vortex. We will pay particular
attention to the correlation between the  chirality of the
worldsheet and four-dimensional fermions.

\para
In the following we use four-dimensional Weyl fermions
$\psi_\alpha$ and $\bar{\lambda}^{\dot{\alpha}}$ with
$\alpha,\dot{\alpha}=1,2$. The notation is standard Wess and
Bagger fare \cite{wb} with, for example,
$\psi\lambda=\psi^\alpha\lambda_\alpha=\lambda\psi$ and
$\bar{\psi}\bar{\lambda}=\bar{\psi}_{\dot{\alpha}}\bar{\lambda}^{\dot{\alpha}}
=\bar{\lambda}\bar{\psi}$. Indices are raised and lowered with
$\epsilon^{\alpha\beta}=\epsilon^{\dot{\alpha}\dot{\beta}}=i\sigma_2$.
Our signature is mostly minus and we define
$(\sigma^\mu)_{\alpha\dot{\alpha}}=(-1,\sigma^i)$ and
$(\bar{\sigma}^\mu)^{\dot{\alpha}\alpha}=(-1,-\sigma^i)$.

\para
The ${\cal N}=2$ vector multiplet in four dimensions contains two
Weyl fermions, $\lambda$ and $\eta$, each transforming in the
adjoint representation of the $U(N_c)$ gauge group. The fermion
$\lambda$ lives in the ${\cal N}=1$ vector multiplet while $\eta$
lives in the adjoint chiral multiplet $A$. Each hypermultiplet
also contains two Weyl fermions, $\psi$ and $\tilde{\psi}$. These
live in $Q$ and $\tilde{Q}$, and transform in the ${\bf N}_c$ and
$\bar{\bf N}_c$ representations respectively. The Dirac equations
in the ${\cal N}=2$ theory are
%
%
\be
-\frac{i}{e^2}\Dbarslash\lambda-\frac{i\sqrt{2}}{e^2}[\bar{\eta},A]+i\sqrt{2}Q_i\bar{\psi}_i-
i\sqrt{2}\bar{\tilde{\psi}}_i\tilde{Q}_i&=&0 \nn\\
-\frac{i}{e^2}\Dbarslash\eta-\frac{i\sqrt{2}}{e^2}
[A,\bar{\lambda}]-\sqrt{2}\tilde{Q}^\dagger_i\bar{\psi}_i-
\sqrt{2}\bar{\tilde{\psi}}_i{Q}_i^{\dagger}&=&0 \label{diracs}\\
-i\Dbarslash\psi_i+i\sqrt{2}\bar{\lambda}Q_i-\sqrt{2}A^\dagger\bar{\tilde{\psi}}_i
-\sqrt{2}\bar{\eta}\tilde{Q}_i^\dagger &=&0\nn\\
-i\Dbarslash\tilde{\psi}_i-i\sqrt{2}\tilde{Q}_i\bar{\lambda}-\sqrt{2}\bar{\psi}_i
A^\dagger -\sqrt{2}Q_i^{\dagger}\bar{\eta}&=&0. \nn\ee
We wish to study these equations in the background of the vortex.
Here they simplify considerably since we have $A=\tilde{Q}_i=0$.
The equations decouple into two pairs: the first set of equations
are for $\lambda$ and $\bar{\psi}_i$
\be -\frac{i}{e^2}\Dbarslash\lambda+i\sqrt{2}Q_i\bar{\psi}_i=0\ \
\ &{\rm and}&\ \ \ \
-i\Dslash\bar{\psi}_i-i\sqrt{2}Q_i^{\dagger}\lambda=0.
\label{lambdap}\ee
The second set of equations are for $\eta$ and
$\bar{\tilde{\psi}}_i$,
\be
-\frac{i}{e^2}\Dbarslash\eta-\sqrt{2}\bar{\tilde{\psi}}_iQ_i^{\dagger}=0\
\ \ &{\rm and}&\ \ \ -i\Dslash\bar{\tilde{\psi}}_i-\sqrt{2}\eta
Q_i=0.\label{etap}\ee

\subsubsection{Chirality}

Each  pair of four-dimensional fermions gives rise to a fermi zero
mode on the vortex string of a specific chirality. Since this will
be important in later sections, we dwell on the point a little
here. The first step is to see which components of the spinors can
turn on in the background of a vortex or anti-vortex. We will need
the following identities,
\be
\Dslash_{\alpha\dot{\alpha}}\equiv(\sigma^\mu)_{\alpha\dot{\alpha}}
{\cal D}_\mu=2\left(\begin{array}{cc} -{\cal D}_- & {\cal D}_z \\
{\cal D}_{\bar{z}} & -{\cal D}_+\end{array}\right) \ \ {\rm and}\
\ \Dbarslash^{\dot{\alpha}\alpha}
\equiv(\bar{\sigma}^\mu)^{\dot{\alpha}\alpha}
{\cal D}_\mu=-2\left(\begin{array}{cc} {\cal D}_+ & {\cal D}_z \\
{\cal D}_{\bar{z}} & {\cal D}_-\end{array}\right)\label{ds}\ \ \ \
\ \ \  \ee
where ${\cal D}_\pm=\ft12({\cal D}_0\pm{\cal D}_3)$ and ${\cal
D}_z=\ft12({\cal D}_1-i{\cal D}_2)$ and ${\cal
D}_{\bar{z}}=\ft12({\cal D}_1+i{\cal D}_2)$. Our strings are
static and oriented in the $x^3$ direction, so in searching for
zero modes of the Dirac equation in the presence of a vortex we
may initially set ${\cal D}_\pm=0$. We decompose the spinors as
$(\lambda_1,\lambda_2)=(\lambda_-,\lambda_+)$ and
$(\lambda^1,\lambda^2)=(\lambda^-,\lambda^+)$ so that, with our
raising and lowering conventions, $\lambda^+=-\lambda_-$ and
$\lambda^-=\lambda_+$. To see which components turn on in the
background of the vortex, we act on the first and second equations
in \eqn{lambdap} with $\Dslash$ and $\Dbarslash$ respectively.
Making use of the vortex equation ${\cal D}_{{z}}Q_i=0$, we find
\be &&(-\frac{4}{e^2}{\cal D}_z{\cal
D}_{\bar{z}}+2Q_iQ_i^{\dagger})\lambda_- =0\nn\\&&
(-\frac{4}{e^2}{\cal D}_{\bar{z}}{\cal
D}_z+2Q_iQ_i^{\dagger})\lambda_+ - \sqrt{2}({\cal
D}_{\bar{z}}Q_i)\,\bar{\psi}_{+i}=0 \ee
and
\be &&(-{\cal D}_z{\cal D}_{\bar{z}}\delta_{ij}
+2Q_iQ_j^{\dagger})\bar{\psi}_{+j} - \sqrt{2}({\cal D}_z
Q_i^{\dagger})\, \lambda_+ =0\nn\\
&&(-{\cal D}_{\bar{z}}{\cal D}_z\delta_{ij}+2Q_i^{\dagger}
Q_j)\bar{\psi}_{-j}  =0. \ee
%
%
%
%
The operators appearing in the equations for $\lambda_-$ and
$\bar{\psi}_{-i}$ are positive definite: these components can have
no zero modes. All zero modes live in the components $\lambda_+$
and $\bar{\psi}_{+i}$.

\para
To see how this correlates with the chirality of the worldsheet
fermions, we now allow these zero modes to vary along the string
so that $\lambda_+=\lambda_+(x^0,x^3)$ and
$\bar{\psi}_{+i}=\bar{\psi}_{+i}(x^0,x^3)$. Plugging this ansatz
back into the Dirac equation, including now the derivatives ${\cal
D}_\pm$ in \eqn{ds}, we find the equations of motion
${\partial}_-\lambda_+=\partial_-\bar{\psi}_{+i}=0$. We call these
fermions {\it right movers}.

\para
Repeating this analysis for Dirac equations \eqn{etap}, we find
that $\eta_-$ and $\bar{\tilde{\psi}}_{-i}$ both carry zero modes
in the background of the vortex. They are {\it left movers} on the
string worldsheet\footnote{In the background of an anti-vortex,
with  ${\cal D}_{\bar{z}}Q_i=0$, the chirality of the fermi zero
modes is reversed, so that $(\lambda, \bar{\psi})$ donate left
movers, while $(\eta,\bar{\tilde{\psi}})$ donate right movers.}.

\subsubsection{Fermi Zero Modes}

From the previous analysis, we learn that the right moving fermi
zero modes solve
\be \sqrt{2}{\cal D}_z\lambda_+ &=& -e^2\,Q_i\bar{\psi}_{+i} \nn\\
\sqrt{2}{\cal D}_{\bar{z}} \bar{\psi}_{+i}&=&-
Q_i^\dagger\lambda_+\ee
while the equations for the left moving fermi zero modes solve
\be \sqrt{2}i{\cal D}_{\bar{z}} \eta_- &=& -e^2
\bar{\tilde{\psi}}_{-i} Q_i^\dagger \nn\\
\sqrt{2}i{\cal D}_z \bar{\tilde{\psi}}_{-i} &=& \eta_-Q_i.\ee
Each of these pairs of equations is the same as the equations for
bosonic zero modes \eqn{bogzero} that are derived by linearizing
the vortex equations and imposing a gauge fixing constraint. The
relationship between the bosonic and fermionic  zero modes is
given by
\be \lambda_+\leftrightarrow \delta A_{\bar{z}}\ \ &{\rm  and}& \
\
\sqrt{2}\bar{\psi}_{+i}\leftrightarrow -i\delta Q_i^\dagger \nn\\
\eta_-\leftrightarrow \delta A_z \ \ &{\rm and}& \ \
\sqrt{2}\bar{\tilde{\psi}}_{-i}\leftrightarrow -\delta
Q_i.\label{relzero}\ee
This mapping between the zero mode profiles is a consequence of
the preserved supersymmetry in the background of the vortex. Using
this, it is trivial to derive the explicit zero modes in the case
of a single $k=1$ vortex.

\subsubsection*{Goldstino Modes}

The bosonic translational modes were given in \eqn{transzero}.
Their fermionic counterparts are
\be \lambda_+=F_{z\bar{z}}\bar{\chi}_+ \ \ &{\rm and}& \ \
\bar{\psi}_{+i}=-\frac{i}{\sqrt{2}} {\cal D}_{z}Q_i^\dagger
\bar{\chi}_+
\nn\\
\eta_-=F_{\bar{z}z}\chi_- \ \ &{\rm and}& \ \
\bar{\tilde{\psi}}_{-i}=-\frac{1}{\sqrt{2}} {\cal D}_{\bar{z}}Q_i
\chi_- .\label{goldy}\ee
Both of these are  Goldstino modes, arising from acting on the
bosonic vortex profile \eqn{solutions} with the two broken
supersymmetries parameterized by $\chi_\pm$.  The above formulae
hold for arbitrary $k$; if we restrict to the explicit $k=1$
solution, we may write these in terms of the vortex profile
function $q(\rho)$,
\be  (\lambda_+)^a_{\ b}= \frac{ie^2v^2}{2r}
(q^2-1)\phi^a\bar{\phi}_i\bar{\chi}_+ \ \  &{\rm and}& \ \
\bar{\psi}^A_{+i}=-\frac{i\sqrt{2}v}{r}(\partial_zq)\,{\phi^a\bar{\phi}_b}
\bar{\chi}_+
\nn\\
\eta_-= -\frac{ie^2v^2}{2r} (q^2-1)\phi^a\bar{\phi}_b\chi_-
 \ \ &{\rm and}& \ \ \bar{\tilde{\psi}}^a_{-i}=- \frac{\sqrt{2}v}{r}
 (\partial_{\bar{z}}q)\,{\phi^a\bar{\phi}_i}
\chi_- .\label{goldy2}\ee

\subsubsection*{Super-Orientation Modes}

The superpartners of the orientational modes are equally easy to
write down. Given the bosonic zero modes \eqn{orientzero}, we have
\be (\lambda_+)^a_{\ b}=2i (\partial_{\bar{z}}q)\,
\phi^a\bar{\xi}_{+b}
 \ \ &{\rm and}& \ \
 \bar{\psi}^a_{+i}=-\frac{iv}{\sqrt{2}}(q^2-1)\phi_i\bar{\xi}^a_{+}
\nn\\
(\eta_-)^a_{\ b}= -2i(\partial_z q)\, {\xi}^a_-\bar{\phi}_b \ \
&{\rm and}& \ \ \bar{\tilde{\psi}}^a_{-i}=
-\frac{v}{\sqrt{2}}(q^2-1) \xi_-^a \bar{\phi}_i .\label{somodes}\ee
It is clear from these expressions, that the redundancy \eqn{iden}
which acts among the $\phi_i$ orientational coordinates, must also
act on the superpartners $\xi_{\pm i}$, so that
\be \phi_i\rightarrow e^{i\alpha}\phi_i\ \ \ {\rm and} \ \ \
\xi_i\rightarrow e^{i\alpha}\xi_i .\label{iden2}\ee
Moreover, the fact that there do not exist orientational
coordinates in the $N=1$ Abelian theory means that we must impose
a constraint on the $\xi_{\pm i}$, namely
\be \sum_{i=1}^{N_c}\bar{\phi}^i\xi_{\pm i}=0 .\label{fermcons}\ee

\subsection{Supersymmetric Dynamics}

The low-energy dynamics of the vortex string arises by promoting
the collective coordinates $z$, $\chi_\pm$, $\phi_i$ and $\xi_{\pm
i}$ to dynamical fields on the string worldsheet, depending on
$y^0\equiv x^0$ and $y^1\equiv x^3$. The fact that the vortices
are BPS, preserving 1/2 of the ${\cal N}=2$ four-dimensional
supersymmetry, ensures that the resulting worldsheet dynamics is
invariant under  ${\cal N}=(2,2)$ supersymmetry. Indeed, the
various bosonic and fermionic collective coordinates are easily
packaged into ${\cal N}=(2,2)$ superfields. The translational mode
$z$ and the two Goldstino modes $\chi_\pm$ sit in an ${\cal
N}=(2,2)$ chiral multiplet $Z$. Our notation is
standard\footnote{The one deviation from standard notation is to
label the complex auxiliary fields in each chiral multiplet as
$G$. This distinguishes them from the auxiliary $F$ fields in
four-dimensions.} and follows, for example, \cite{phases}
\be
Z=z+\theta^+\chi_++\theta^-\chi_-+\theta^+\theta^-G_Z+\ldots\ .\ee
Similarly, the orientation modes $\phi_i$ and their superpartners
$\xi_{\pm i}$ also sit in $(2,2)$ chiral multiplets,
\be
\Phi_i=\phi_i+\theta^+\xi_{+i}+\theta^-\xi_{-i}+\theta^+\theta^-G_i+\ldots\ .
\ee
The two constraints $\bar{\phi}^i\phi_i=r$, and
$\bar{\phi}^i\xi_{\pm i}=0$, together with the identification
\eqn{iden2}, are imposed on the worldsheet theory by introducing
an auxiliary ${\cal N}=(2,2)$ vector multiplet which, in
Wess-Zumino gauge, has components
\be
U&=&-\theta^-\bar{\theta}^-(u_0-u_1)+\theta^+\bar{\theta}^+(u_0+u_1)
- \theta^-\bar{\theta}^+\sigma -
\theta^+\bar{\theta}^-\bar{\sigma}
\\ && + \sqrt{2}{i}\theta^i\theta^+(\bar{}\theta}^-\bar{\zeta}_-
+\bar{\theta}^+\bar{\zeta}_+)+\sqrt{2}{i\bar{\theta}^+\bar{\theta}^-
(\theta^-\zeta_-+\theta^+\zeta_+)+2\theta^-\theta^+
\bar{\theta}^+\bar{\theta}^- D .\nn\ee
The two dimensional field strength
$u_{01}=\partial_0u_1-\partial_1 u_0$ is naturally housed in a
twisted chiral multiplet, defined by $\Sigma =
\bar{D}_+D_-U/\sqrt{2}$, with component expansion
\be \Sigma = \sigma -
i\sqrt{2}\theta^+\bar{\zeta}_+-i\sqrt{2}\bar{\theta}^-\zeta_-+\sqrt{2}
\theta^+\bar{\theta}^-(D-iu_{01})+\ldots\ .\label{twistedc}\ee
The fields $\sigma$, $\zeta_\pm$ and $D$ are all auxiliary. Their
role will become clear shortly.

\para
With the exception of a single integration constant $t$, the
dynamics of a $k=1$ vortex string is fixed entirely by the
symmetries of the theory. In particular, the $SU(N_c)_{\rm diag}$
symmetry of \eqn{breaking} descends to an $SU(N_c)$ global
symmetry on the worldsheet, under which the $\Phi_i$ transform in
the fundamental ${\bf N_c}$ representation. The resulting
dynamics is given by
\be {\cal L}_{\rm vortex} = \int d^4\theta\ T\bar{Z}Z+
\sum_{i=1}^{N_c}\bar{\Phi}_ie^{2U}\Phi_i + \frac{it}{2\sqrt{2}}\int d\theta^+
d\bar{\theta}^-\ \Sigma\ee
where $T=2\pi v^2$ is the tension of the vortex, while
\be t = ir+\frac{\theta}{2\pi}\ee
is the integration constant that needs to be fixed, and plays the
role of a complexified worldsheet FI parameter. After integrating
out the auxiliary fields $G_Z$ and $G_i$, the purely bosonic part
of the worldsheet Lagrangian reads,
\be {\cal L}_{\rm bose}=T\,|\partial_m z|^2 +
\sum_{i=1}^{N_c}\left(|{\cal D}_m\phi^i|^2 -
2|\sigma|^2|\phi_i|^2\right)
+D(\sum_{i=1}^{N_c}|\phi_i|^2-r)+\frac{\theta}{2\pi}u_{01}.\ \ \ \
\ \ee
Here the $\phi_i$ fields carry charge $+1$ under the gauge
symmetry, with ${\cal D}\phi_i=\partial\phi_i - iu\phi_i$.
Dividing out by this symmetry imposes the identification
\eqn{iden}: $\phi_i\rightarrow e^{i\alpha}\phi_i$. Meanwhile, the
$D$-field in this Lagrangian plays the role of a Lagrange
multiplier, imposing the condition \eqn{phiphi}:
$\sum_i|\phi_i|^2=r$. The value of $r$ is fixed by the requirement
that the kinetic terms for $\phi_i$ are canonical. One finds the
result,
\be r=\frac{2\pi}{e^2}.\ee
This result was first shown using a brane construction in
\cite{vib}, and later re-derived by explicitly computing the
overlap of zero modes\footnote{This follows by taking the time
dependent ansatz for the orientational modes: ${\cal
D}_tQ_i=\delta Q_i$ and $F_{0z}=\delta A_z$, with
$\delta\phi_i=\dot{\phi}_i$. Inserting this into the four
dimensional kinetic terms gives,
\be \int dx^1dx^2 \left(\frac{1}{2e^2}F_{0i}^2+|{\cal
D}_tQ_i|^2\right) = \int dx^1dx^2 \left(\frac{1}{e^2}|{\cal
D}_iq|^2 +v^2(q-1)^2(q+1)^2\right)\frac{|{\cal D}_t\phi_i|^2}{r} =
\frac{2\pi}{e^2} \frac{|{\cal D}_t\phi_i|^2}{r}\nn\ee where the
integral is recognized as the same one that appears in computing
the vortex tension.} in \cite{others}. Similarly, it can be shown
that the four-dimensional $\theta$-angle descends to a worldsheet
$\theta$-angle \cite{vstring,others}. The end result is that
worldsheet complexified FI parameter $t$ is identified with the
bulk complexified gauge coupling $\tau$ \eqn{tau}:
\be t = \tau\ .\ee
We now turn to the fermionic part of the worldsheet Lagrangian,
given by
\be L_{\rm fermi} &=& 2iT\left(\bar{\chi}_-\partial_+\chi_-
+\bar{\chi}_+\partial_-\chi_+\right) +
2i\sum_{i=1}^{N_c}\left(\bar{\xi}_{-i}{\cal D}_+\xi_{-i} +
\bar{\xi}_{+i}{\cal D}_-\xi_{+i}\right)\\ &&
-\sqrt{2}\sum_{i=1}^{N_c}\left(
\bar{\sigma}\bar{\xi}_{+i}\xi_{-i}+\sigma \bar{\xi}_{-i}\xi_{+i}
+\bar{\phi}_i(\xi_{-i}\zeta_+-\xi_{+i}\zeta_-)+\phi_i(\bar{\zeta}_-\bar{\xi}_{+i}
-\bar{\zeta}_+\bar{\xi}_{-i})\right)\nn\ee
The fermions $\xi_{\pm i}$ both have charge $+1$ under the $U(1)$
gauge symmetry, which is now seen to implement the full
identification \eqn{iden2}: $\phi_i\rightarrow e^{i\alpha}\phi_i$
and $\xi_\pm\rightarrow e^{i\alpha}\xi_{\pm}$. The vector
multiplet fermions $\zeta_\pm$ have no kinetic term and act as
Grassmannian Lagrange multipliers, imposing the constraint
\eqn{fermcons}: $\sum_i\bar{\phi}_i\xi_{\pm i}=0$. Finally, the
role of $\sigma$ is to mediate a four-fermi interaction for the
super-orientation modes. Upon integrating out $\sigma$, we have
\be {\cal L}_{\rm 4-fermi} =
-2|\sigma|^2|\phi_i|^2-\sqrt{2}\bar{\sigma}\bar{\xi}_{+i}\xi_{-i}-\sqrt{2}
\sigma\bar{\xi}_{-i} \xi_{+i} =
-\frac{|\bar{\xi}_{-i}\xi_{+i}|^2}{r}.\label{224fermi}\ee
Four-fermi terms of this kind are typical for soliton dynamics in
supersymmetric theories. We pause here to review how they arise.
In deriving the Dirac equations \eqn{lambdap} and \eqn{etap} we
set $A=\tilde{Q}_i=0$. This is valid in the background of the
bosonic vortex. However, it is no longer true in the presence of
fermions since fermi bilinears act as a source for these fields.
For example, the Yukawa couplings involving $A^\dagger$ contribute
to the equation of motion,
\be {\cal D}^2A +
{i\sqrt{2}}[\lambda,\eta]-\sqrt{2}e^2\bar{\tilde{\psi}}_i\tilde{\psi}_i
+e^2\{Q_iQ_i^\dagger + \tilde{Q}_i^\dagger\tilde{Q}_i, A\} =0
\label{yukyuk}\ee
%
The solution to this equation then feeds back into the Dirac
equations \eqn{diracs} and must be solved iteratively, order by
order in the number of Grassmannian collective coordinates. This
is a finite, but somewhat complicated procedure (see \cite{stefan}
for a simple quantum mechanical model where it may be carried
through to completion). Thankfully, the end result \eqn{224fermi}
is dictated by supersymmetry.

\subsubsection{Symmetries}
\label{symmetries}

The four-dimensional ${\cal N}=2$ theory has two $U(1)$
R-symmetries\footnote{In the absence of the FI parameter $v^2$,
the theory has an $SU(2)_R$ symmetry, under which $(\lambda,\eta)$
and $(Q,\tilde{Q}^\dagger)$ both transform as doublets. The FI
parameter breaks $SU(2)_R\rightarrow U(1)_V$.} that we will call
$U(1)_R$ and $U(1)_V$. The charges of the various fields under
$U(1)_R\times U(1)_V$ are listed in the table.
\begin{center}\begin{tabular}{c|ccccccc} & $A$ & $\lambda$ & $\eta$ &
$Q$ & $\tilde{Q}$ & $\psi$
& $\tilde{\psi}$ \\ \hline $U(1)_R$ & 2 & 1 & 1 & 0 & 0 & -1 & -1 \\
$U(1)_V$ & 0 & 1 & -1 & 1 & 1 & 0 & 0\ .
\end{tabular}\end{center}
Both of these symmetries descend to the vortex worldsheet, where
they appear as the two R-symmetries of the ${\cal N}=(2,2)$
superalgebra. The action on the fermionic collective coordinates
of the vortex can  can be read directly from  \eqn{relzero}. We
have
\begin{center}\begin{tabular}{c|ccccccccc} & $\sigma$ & $\zeta_+$
& $\zeta_-$ & $\phi$ & $\xi_+$ & $\xi_-$ & $z$ & $\chi_+$ &
$\chi_-$ \\ \hline $U(1)_R$ & 2 & -1 & 1 & 0 & -1 & 1 & 0 & -1 & 1 \\
$U(1)_V$ & 0 & 1 & 1 & 0 & -1 & -1 & 0 & -1 & -1
\\ $U(1)_Z$ & 0 & 1 & 1 & 0 & -1 & -1 & 2 & 1 & 1\ .
\end{tabular}\end{center}
The $U(1)_R$ symmetry is axial; it suffers an anomaly in the
quantum theory of the vortex (as, indeed, does the $U(1)_R$ in
four-dimensions). In contrast $U(1)_V$ is a vector R-symmetry on
the worldsheet.

\para
The vortex theory also includes a further global $U(1)_Z$
symmetry, which arises from rotating the vortex string in the
$z=x^1+ix^2$ plane. The charges of the worldsheet fields under
$U(1)_Z$ are listed in the table and follow from \eqn{goldy} and
\eqn{somodes}. There exists a suitable linear combination of
$U(1)_Z$ and $U(1)_V$ which simply rotates the phase of the chiral
multiplet $Z$, leaving all other fields invariant.

\para
There are other, translational, symmetries of the worldsheet
theory that reflect the fact that $z$ and $\chi_\pm$ are all
Goldstone modes, arising from broken translation and supersymmetry
invariance respectively. In both cases, this ensures they have
only derivative couplings. In particular, it is the existence of
these symmetries that prevents the Goldstino modes $\chi_\pm$ from
appearing in the four-fermi term \eqn{224fermi}.

\subsection{${\cal N}=2$ Preserving Deformations}

So far we have described vortices in only the simplest ${\cal
N}=2$ theory with $N_f=N_c$. There are a number of ways to deform
and augment our theory that preserve ${\cal N}=2$ supersymmetry.
Here we list them and describe their effect on the worldsheet. We
postpone until Section 4 a discussion of deformations that break
the four dimensional supersymmetry to ${\cal N}=1$.

\subsubsection{Adding Masses}

The simplest deformation of our theory that preserves ${\cal N}=2$
supersymmetry is to add a complex mass parameter $m_i$ for each
hypermultiplet. The superpotential \eqn{4dn2} now becomes
\be {\cal W}_{{\cal
N}=2}=\sqrt{2}\sum_{i=1}^{N_f}\tilde{Q}_i(A-m_i)Q_i .\ee
The vacuum \eqn{vac1} survives only if we turn on the adjoint
scalar field $A$ to cancel the  F-term contributions,
\be Q^a_{\ i}=v\delta^a_{\ i}\ \ \ ,\ \ \ \tilde{Q}_i=0\ \ \ \ , \
\ \ A={\rm diag}(m_1,\ldots,m_{N_c}) .\label{vac2}\ee
The vortex moduli space does not fare well under this deformation.
It can be simply shown that the masses $m_i$ lift the internal
${\bf CP}^{N_c-1}$ vortex moduli space, leaving behind $N_c$
distinct, isolated vortex solutions, each of which carries
magnetic flux in a different diagonal $U(1)$ subgroup of the
$U(N_c)$ gauge group, supported by a different $Q_i$ winding at
infinity.

\para
It was shown in \cite{memono,sy,vstring} that the 4d masses $m_i$
induce ``twisted masses" \cite{hh} for the fields on the vortex
worldsheet. In the language of ${\cal N}=(2,2)$ superfields, this
deformation replaces the standard kinetic terms for $\Phi_i$
by
\be \sum_{i=1}^{N_c}\bar{\Phi}_i e^{2U}\Phi_i \longrightarrow
\sum_{i=1}^{N_c}\ \bar{\Phi}_i
\exp\left(2U-2\theta^-\bar{\theta}^+m_i-2\theta^+\bar{\theta}^-m_i^\dagger\right)
\Phi_i.\ee
In terms of components, the vortex theory (neglecting for now the
$Z$ multiplet whose dynamics remains unchanged) becomes,
\be {\cal L}_{\rm vortex} &=& \sum_{i=1}^{N_c}\left(|{\cal D}_m\phi_i|^2
+|F_i|^2- 2|\sigma-m_i|^2|\phi_i|^2\right)
+D(\sum_{i=1}^{N_c}|\phi_i|^2-r)+\frac{\theta}{2\pi}u_{01} \nn\\
&& +\sum_{i=1}^{N_c} 2i\left(\bar{\xi}_{-i}{\cal D}_+\xi_{-i} +
\bar{\xi}_{+i}{\cal D}_-\xi_{+i}\right) -\sqrt{2}
(\bar{\sigma}-\bar{m}_i)\bar{\xi}_{+i}\xi_{-i}
-\sqrt{2}(\sigma-m_i) \bar{\xi}_{-i}\xi_{+i} \nn\\ &&
-i\sqrt{2}\sum_{i=1}^{N_c}\Big(
\bar{\phi}_i(\xi_{-i}\zeta_+-\xi_{+i}\zeta_-)+\phi_i(\bar{\zeta}_-\bar{\xi}_{+i}
-\bar{\zeta}_+\bar{\xi}_{-i})\Big).\ee
Note that the masses $m_i$ break the $U(1)_R$ symmetry, both in
four-dimensions and on the vortex worldsheet. The twisted masses
on the worldsheet have the desired effect of lifting the ${\bf
CP}^{N_c-1}$ moduli space of the vortex theory, leaving behind
$N_c$ isolated vacua given by
\be |\phi_i|^2 = r\delta_{ij}\ \ \ ,\ \ \ \sigma = m_j\ \ \ \
j=1,\ldots, N_c .\ee
These different vacua of the worldsheet theory are identified with
the different vortex solutions in four-dimensions. Kinks
interpolating between these vacua on the worldsheet correspond to
magnetic monopoles in four-dimensions, confined to lie on the
vortex string by the Meissner effect \cite{memono}.

\subsubsection{Adding Flavors}
\label{add}

We now consider the theory with $N_f>N_c$ fundamental
hypermultiplets. The D-term and F-term vacuum conditions in
four-dimensions  read
\be \sum_{i=1}^{N_f}Q_iQ_i^\dagger - \tilde{Q}_i^\dagger
\tilde{Q}_i = v^2\ \ \ {\rm and}\ \ \ \sum_{i=1}^{N_f}
Q_i\tilde{Q}_i= 0 .\label{higgsbranch}\ee
When $m_i=0$, there are no further conditions, and there is a
$2N_c(N_f-N_c)$ dimensional Higgs branch of the theory. For the
purposes of this section, we place ourselves in the  particular
vacuum $\tilde{Q}_i=0$ and
\be Q^a_{\ i}=v\delta^a_{\ i} \ \ \ {a,i=1,\ldots N_c}
\label{0vac}\ee
with $Q_i=0$ for $i=N_c+1,\ldots, N_f$. This is to be supplemented
by $A=0$. If we now turn on masses for the hypermultiplets, the
vacuum \eqn{0vac} survives, with $A={\rm
diag}(m_1,\ldots,m_{N_c})$.

\para
Vortices in theories with $N_f>N_c$ have a rather different
character than those in the $N_f=N_c$ theory. The most noticeable
difference is that they gain extra bosonic collective coordinates,
among them a scale size. These additional collective coordinates
are non-normalizable when $m_i=0$ \cite{ward,ls} but become
normalizable when finite masses $m_i$ are turned on for
$i=N_c+1,\ldots, N_f$ \cite{semi}. Vortices of this kind are
sometimes referred to as semi-local vortices: a review of these
objects in Abelian theories can be found in \cite{semi2}, while a
detailed discussion in non-Abelian theories was given in
\cite{semi}.

\para
An effective dynamics for the vortex worldsheet in theories with
$N_f>N_c$ was proposed in \cite{vib}, based on a D-brane
construction. It is once again an ${\cal N}=(2,2)$ supersymmetric
$U(1)$ gauge theory, now  with $N_c$ chiral multiplets $\Psi_i$ of
charge $+1$ and a further $(N_f-N_c)$ chiral multiplets
$\tilde{\Psi}_j$ of charge $-1$. The D-term for this theory reads
\be D= \sum_{i=1}^{N_c} |\phi_i|^2 -
\sum_{j=1}^{N_f-N_f}|\tilde{\phi}_j|^2-r = 0 \label{tilphi}\ee
which, together with the gauge action $\phi_i\rightarrow
e^{i\alpha}\phi_i$ and $\tilde{\phi}_i\rightarrow
e^{-i\alpha}\tilde{\phi}_i$, defines the Higgs branch of the
vortex theory. This Higgs branch  is conjectured to coincide the
vortex moduli space. As in the previous section, assigning complex
masses $m_i$ to the four dimensional hypermultiplets induces
twisted masses $m_i$, $i=1,\ldots,N_c$ for the $\Phi_i$ fields,
and twisted masses $\tilde{m}_j=m_{j+N_c}$, $j=1,\ldots, N_f-N_c$
for $\tilde{\Phi}_j$.

\para
The presence of the negatively charged fields $\tilde{\phi}_j$
means that the moduli space \eqn{tilphi} is now non-compact,
corresponding to the scaling mode of the vortex. Note however that
the natural metric on the Higgs branch does not coincide with the
natural metric on the vortex moduli space. In particular, the
non-normalizability of the scaling modes as $m_i\rightarrow 0$ is
not reproduced in this model.  Nonetheless, it has been shown that
the vortex theory \eqn{tilphi} does indeed correctly capture the
quantum dynamics of the vortex string \cite{vstring,scvs}.

\subsubsection*{Higgs Expectation Values}

When $N_f>N_c$ and $m_i=0$, the vacuum conditions
\eqn{higgsbranch} in the four-dimensional theory have a moduli
space of solutions. We may ask what happens to the vortex string
as we change the expectation values of $Q_i$ and $\tilde{Q}_i$
such that \eqn{higgsbranch} remains satisfied. The answer to this
question was given in \cite{scvs}: turning on expectation values
for $\tilde{Q}_i$ induces a superpotential on the vortex string
worldsheet. For completeness, we briefly describe this deformation
here.

\para
First some notation: define the gauge invariant meson operator
\be M_i^j\equiv \tilde{Q}^j Q_i\label{mij} .\ee
It is not hard to show that in four-dimensional vacua for which
$M\neq 0$, the space of BPS vortex solutions is greatly reduced.
The key point is that a vacuum expectation value for $\tilde{Q}$
does not allow a BPS vortex to live in the associated part of the
gauge group. This follows from the mathematical fact that there is
no holomorphic line bundle of negative degree. In a more physical
language, a direct analysis of the vortex equations reveals that
BPS vortices do not exist in $U(1)$ theories if both negatively
and positively charged fields gain an expectation value
\cite{penin,davis}. The upshot of this is that the vortex moduli
space is partly lifted in four dimensional vacua for which $M\neq
0$. It was shown in \cite{scvs} that this effect is captured on
the vortex worldsheet by the introduction of a superpotential of
the form,
\be {\cal W}_{(2,2)}\sim \sum_{i=1}^{N_f} \sum_{j=1}^{N_f-N_c}
M^{j+N_c}_i \,\tilde{\Phi}_j\Phi^i .\ee

\subsection{Multiple Vortices}

\EPSFIGURE{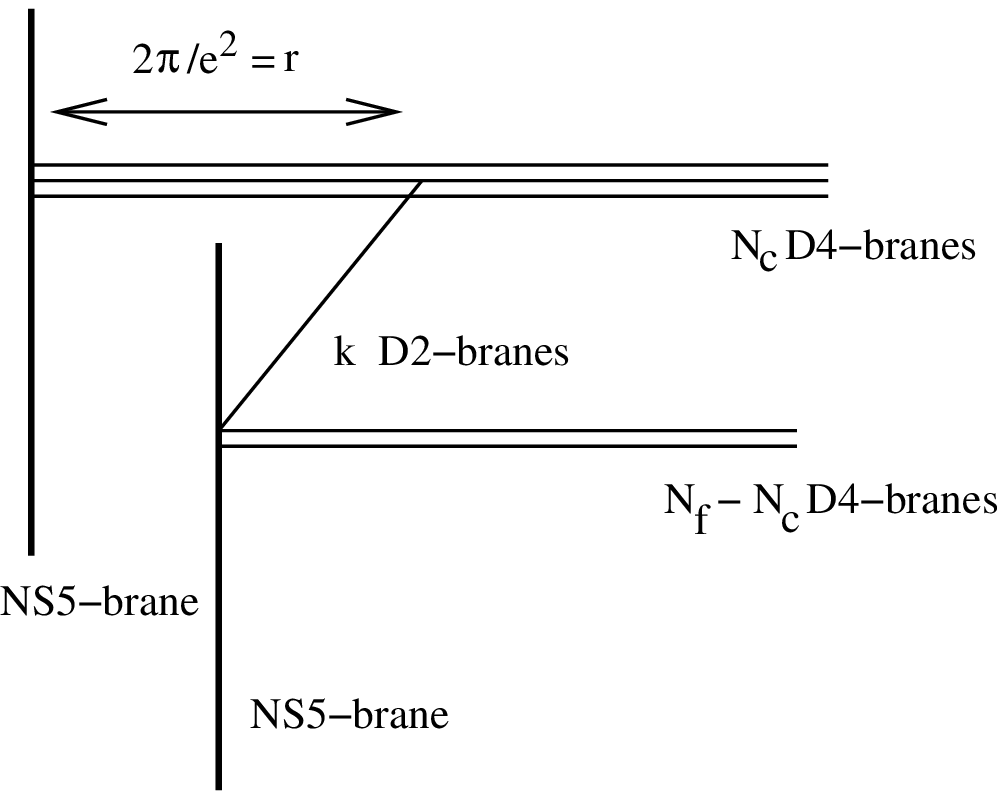,height=160pt}{}
\noindent So far we have focussed on the theory for a single
vortex string for which, at least in the case $N_f=N_c$,
symmetries are sufficient to dictate the dynamics. In \cite{vib},
a D-brane construction was used to derive a worldsheet theory
which describes the interactions of $k>1$ parallel vortex strings.
The D-brane construction starts with the usual  Hanany-Witten
set-up for ${\cal N}=2$ four-dimensional gauge theories
\cite{hanwit,w}, consisting of D4-branes attached to parallel
NS5-branes. Separating the NS5-branes in the direction out of the
page induces the FI parameter $v^2$. The vortex strings arise as
stretched D2-branes as shown in figure 2. The worldvolume theory
of $k$ vortex strings is given by an ${\cal N}=(2,2)$  $U(k)$
non-Abelian gauge theory with matter content,
\para

\be \mbox{$U(k)$ Vector Multiplet $U$} &+& \mbox{Adjoint Chiral
Multiplet $Z$}\nn\\ &+& \mbox{$N_c$ Fundamental Chiral Multiplets
$\Phi_i$} \nn\\ &+& \mbox{$N_f-N_c$ Anti-Fundamental Chiral
Multiplets $\tilde{\Phi}_j$} . \nn\ee
The complexified worldsheet FI parameter is again equated to the
4d complexified gauge coupling, $t=\tau$, or
\be ir+\frac{\theta_{2d}}{2\pi}=\frac{2\pi
i}{e^2}+\frac{\theta_{4d}}{2\pi}\label{fi} . \ee
The D-term condition for the worldsheet theory is now $u(k)$
valued and is given by,
\be \sum_{i=1}^{N_c}\phi_i\phi_i^\dagger
-\sum_{j=1}^{N_f-N_c}\tilde{\phi}_j^\dagger\tilde{\phi}_j+T[z,z^\dagger]=r\,1_k .
\label{vortexd}\ee
This provides $k^2$ constraints on the $2k(N_f+k)$ degrees of
freedom in $\phi_i$, $\tilde{\phi}_j$ and $z$. After dividing by
$U(k)$ gauge transformations, we are left with a $2kN_f$
dimensional manifold which defines the target space for the vortex
string sigma-model. This was conjectured in \cite{vib} to coincide
with $2kN_f$ dimensional vortex moduli space. This quotient
construction has subsequently been derived from a direct analysis
of the non-Abelian vortex equations \cite{mmatrix,mmreview}.

\para
%
The $4kN_c$ fermionic zero modes of $k$ parallel vortex strings
live in the $U(k)$ adjoint valued $\chi_\pm$ and the fundamental
$\xi_{\pm i}$, subject to the $2k^2$ complex constraints arising
from the auxiliary fermions $\zeta_\pm$,
\be \sum_i\phi_i\bar{\xi}_{\pm i}+[z,\bar{\chi}_\pm]=0 .\ee
In the case of a single $k=1$ vortex, these reduce to the
constraints \eqn{fermcons}.

\para
Note that the vacuum moduli space \eqn{vortexd} inherits a metric
from the canonical kinetic terms for $\phi_i$ and $z$. This metric
is known not to agree with the standard Manton metric
\cite{manton,samols} on the vortex moduli space (except in the
special case $k=1$ and $N_f=N_c$ that we described in detail
earlier). This is because the limit in which the $d=1+1$ gauge
theory on the D2-branes decouples from other stringy modes is
different from the limit in which the D2-branes are described as
vortices in the $d=3+1$ dimensional theory on the D4-branes; the
two descriptions hold in different regimes of validity as we vary
the parameters of the brane set-up. Nonetheless, if one is
interested in computing objects protected by supersymmetry
--- such as the classical, or quantum, masses of BPS states in the
vortex theory ---- it should be valid to work with the gauge
linear sigma model. In practice, this claim has been confirmed
only for the $k=1$ theory with $N_f>N_c$. It has also been
confirmed that the intricate topology of the $k=2$ vortex string
moduli space in the $N_f=N_c=2$ is correctly captured by the
gauged linear sigma model \cite{asy,kt,lots}.

\section{${\cal N}=(0,2)$ Supersymmetry}

In the previous section we have worked with both ${\cal N}=1$
superfields in four dimensions, $A$, $Q_i$ and $\tilde{Q}_j$, as
well as ${\cal N}=(2,2)$ superfields in two dimensions, $\Sigma$,
$\Phi_i$ and $\tilde{\Phi}_j$. From now on we will deal with
${\cal N}=(0,2)$ supersymmetry in two dimensions. Since this may
be less familiar to some readers we devote this section to a
review of the structure of ${\cal N}=(0,2)$ superfields \cite{hw,ds} and
their relationship to ${\cal N}=(2,2)$ theories. The presentation
follows \cite{phases} and \cite{abs}.

\subsection{Superfields}

${\cal N}=(0,2)$ supersymmetry is generated by two right-moving,
and no left-moving, supersymmetries. The two chiral supercharges
are $Q_+$ and $\bar{Q}_+$. The $(0,2)$ superspace is parameterized
by the bosonic coordinates $y^\pm=(y^0\pm y^1)$ and their
fermionic partners $\theta^+$ and $\bar{\theta}^+$. The action of
the supersymmetry generators in superspace is given as
\be
Q_+&=&\frac{\partial}{\partial\theta^+}+i\bar{\theta}^+(\partial_0+\partial_1)
\nn\\
\bar{Q}_+&=&-\frac{\partial}{\partial\bar{\theta}^+}-i\theta^+(\partial_0+\partial_1) .
\ee
These commute with the superderivatives,
\be D_+&=&\frac{\partial}{\partial\theta^+}-i\bar{\theta}^+
(\partial_0+\partial_1)
\nn\\
\bar{D}_+&=&-\frac{\partial}{\partial\bar{\theta}^+}+i\theta^+
(\partial_0+\partial_1) \ee
which satisfy $\{D_+,D_+\}=\{\bar{D}_+\bar{D}_+\}=0$ and
$\{D_+,\bar{D}_+\}=2i\partial_+$. We now describe the different
superfields of interest.

\subsubsection*{Gauge Multiplets}

We start with the real, adjoint valued,  gauge multiplet $U$,
which has the component expansion
\be
U=(u_0-u_1)-2i\theta^+\bar{\zeta}_--2i\bar{\theta}^+\zeta_-+2\theta^+\bar{\theta}^+D .
\label{02vec}\ee 
Already we see the chiral nature of the supersymmetry, since only
the combination $u_-=u_0-u_1$ of the two-dimensional gauge field
appears in the superfield, together with a left moving fermion
$\zeta_-$. The scalar field $D$ will be seen to be auxiliary. The
covariant superderivatives are given by
\be {\cal D}_+&=&\frac{\partial}{\partial
\theta^+}-i\bar{\theta}^+({\cal D}_0+{\cal D}_1)
\ \ \ \nn\\ \bar{{\cal D}}_+&=&-\frac{\partial}{\partial
\bar{\theta}^+}+i{\theta}^+({\cal D}_0+{\cal D}_1)\nn\ee
where
\be {\cal D}_0+{\cal D}_1=\partial_0+\partial_1-i(u_0+u_1) \nn\ee
includes the $u_+$ component of the gauge field, but no fermions.
Meanwhile, the gauginos are included in the remaining covariant
superderivative,
\be {\cal D}_0-{\cal D}_1=\partial_0-\partial_1-iU .\ee
The field strength lives naturally in a fermi multiplet (which we
shall define shortly) given by the usual commutator of
derivatives:
\be \Upsilon=[\bar{\cal D}_+,{\cal D}_0-{\cal
D}_1]=-2\Big(\zeta_--i\theta^+(D-iv_{01})-i\theta^+\bar{\theta}^+({\cal
D}_0+{\cal D}_1)\zeta_-\Big) .\ee
Here the field strength is
$u_{01}=\partial_0u_1-\partial_1u_0-i[u_0,u_1]$. The kinetic terms
for the gauge multiplet are then given by integration over all of
superspace $d^2\theta =d\theta^+\,d\bar{\theta}^+$,
\be S_{\rm gauge}&=&\frac{1}{8g^2}\,\Tr\,\int d^2y \,d^2\theta\ \Upsilon^\dagger \Upsilon \nn\\
&=&\frac{1}{g^2}\,\Tr\,\int d^2y\ \Big(\ft12\
u_{01}^2+i\bar{\zeta}_-({\cal D}_0+{\cal D}_1)\zeta_-+D^2\Big)
\ee
but, in fact, will not be required in the following.

\subsubsection*{Chiral Multiplets}

The chiral multiplets of $(0,2)$ theories are bosonic superfields
$\Phi$, living in any representation $R$ of the gauge group. They
satisfy
\be \bar{\cal D}_+\Phi=0 .\ee
Chiral multiplets contain right-moving fermions $\xi_+$, paired
with a complex boson $\phi$. Their component expansion gives
\be
\Phi=\phi+\sqrt{2}\theta^+\xi_+-i\theta^+\bar{\theta}^+(D_0+D_1)\phi
\label{02chi}\ee
where $(D_0+D_1)$ is now the usual bosonic covariant derivative.
The kinetic terms for the chiral multiplet are given by the
action,
\be S_{\rm chiral}&=&-\frac{i}{2}\int d^2y\,d^2\theta\
\bar{\Phi}({\cal D}_0-{\cal D}_1)\Phi \label{chirallag}\\ &=& \int
d^2y\ \left(-|D_\alpha\phi|^2+i\bar{\xi}_+(D_0-D_1)\xi_+ -
i\sqrt{2}\bar{\phi}\zeta_-\xi_+ +
i\sqrt{2}\bar{\xi}_+\bar{\zeta}_-\phi + \bar{\phi}D\phi\right) .
\nn\ee
The scalar field $\phi$ couples to the auxiliary field $D$, to
give rise to the usual D-term (Note that for Abelian theories, if
$\Phi$ has charge $p$ then one should replace $\zeta_-\rightarrow
p\zeta_-$ and $D\rightarrow pD$ in the above action.).

\subsubsection*{Fermi Multiplets}

One novel feature of $(0,2)$ theories that is not shared by the
non-chiral $(2,2)$ theories is the existence of a fermionic
multiplet $\Gamma$, containing only left moving fermions $\chi_-$
and no propagating bosons. Like the chiral multiplets, they can
live in any representation $R$ of the gauge group. The fermi
multiplet satisfies
\be \bar{\cal D}_+\Gamma=\sqrt{2}E\label{e}\ee
where $\bar{\cal D}_+E=0$, which can be solved by taking $E$ to be
a holomorphic function of chiral superfields $E=E(\Phi_i)$. The
fermi multiplet has component expansion
\be \Gamma = \chi_--\sqrt{2}\theta^+ G
-i\theta^+\bar{\theta}^+(D_0+D_1)\chi_--\sqrt{2}\bar{\theta}^+E .
\label{02fer}\ee
Note that the superfield $\Upsilon$ containing the field strength
is of this type, with $\bar{\cal D}_+\Upsilon = 0$. In general,
$E$ itself will also have a $\theta$ expansion,
\be E(\Phi_i)=E(\phi_i)+\sqrt{2}\theta^+\frac{\partial
E}{\partial\phi_i}\xi_{+i}
-i\theta^+\bar{\theta}^+(D_0+D_1)E({\phi_i}) \ee
The kinetic terms for the fermi multiplet are
\be S_{\rm fermi}&=&-\frac{1}{2}\int d^2y\,d^2\theta\
\bar{\Gamma}\Gamma \label{fermilag}\\ &=&
\left(i\bar{\chi}_-(D_0+D_1)\chi_-+|G|^2-|E(\phi_i)|^2
-\bar{\chi}_-\frac{\partial E}{\partial\phi^i}\xi_{+i}
+\bar{\xi}_{+i}\frac{\partial\bar{E}}{\partial\bar{\phi}_i}\chi_-\right)
\nn\ee
We see that the complex scalar $G$ is an auxiliary field, lacking
a kinetic term. Also note that the function $E(\phi)$ appears as a
potential term in the Lagrangian.

\subsection{Superpotentials}

In ${\cal N}=(0,2)$ theories the auxiliary field $G$ lives in a fermi
multiplet $\Gamma$, rather than a chiral multiplet. A
superpotential $J(\Phi_i)$ is a holomorphic function of chiral
superfields and a suitable action may be constructed by
integrating terms of the form $\Gamma J$ over half of superspace.
Most generally we can introduce a superpotential $J^a$ for each
fermi multiplet $\Gamma^a$,
\be S_{J} &=& -\frac{1}{\sqrt{2}}\sum_a\int d^2y\,d\theta^+ \ \
\Gamma_a\left.
J^a(\Phi_i)\right|_{\bar{\theta}^+=0}+\ {\rm h.c.} \nn\\
&=& \sum_{a}\int d^2y\ \
G_aJ^a(\phi_i)+\sum_i\chi_{-a}\frac{\partial
J^a}{\partial\phi_i}\xi_{+i}+\ {\rm h.c.}\ .\label{sj}\ee
This integration over half of superspace yields an ${\cal
N}=(0,2)$ supersymmetric invariant action if and only if
$\bar{D}_+(\Gamma_aJ^a)=0$, which requires
\be \sum_a E_aJ^a=0 \label{ej} .\ee
Of course, the combination $\Gamma_aJ^a$ is also required to be
gauge invariant. An important example of the superpotential is the
Fayet-Iliopoulos and theta term which are packaged in the complex
combination $t=ir+\theta/2\pi$. The interaction can be written as
\be S_{D\theta}&=&\frac{t}{4}\,\Tr\int d^2y d\theta^+\
\left.\Upsilon\right|_{\bar{\theta}^+=0} +{\rm \ h.c.}\nn\\ &=&
\Tr\int d^2y\ (-rD+\frac{\theta}{2\pi}u_{01}) .\label{fipot}\ee

\subsection{${\cal N}=(2,2)$ Decomposition}

It will prove useful for orientation to recall how the more
familiar ${\cal N}=(2,2)$ superfields decompose into their ${\cal
N}=(0,2)$ counterparts. The conventions below are taken from
\cite{phases}.

\para
One can enlarge ${\cal N} =(0,2)$ superspace to ${\cal N}=(2,2)$ superspace through
the addition of two further fermionic components $\theta^-$ and
$\bar{\theta}^-$. The corresponding superderivatives are
\be
D_-=\frac{\partial}{\partial\theta^-}-i\bar{\theta}^-(\partial_0-\partial_1)\
\ \ \ ,\ \ \ \
\bar{D}_-=-\frac{\partial}{\partial\bar{\theta}^-}+i\theta^-(\partial_0-\partial_1).\
\ee
The ${\cal N}=(2,2)$ vector multiplet $V_{(2,2)}$ decomposes into
an ${\cal N}=(0,2)$ vector multiplet $V$ described in \eqn{02vec},
together with an ${\cal N}=(0,2)$ chiral multiplet $\Sigma$. This
chiral multiplet inherits the right moving fermion $\zeta_+$ and
the complex scalar field $\sigma$ contained in $V_{(2,2)}$. It is
most simply described by reduction from the ${\cal N}=(2,2)$
twisted chiral multiplet containing the field strength
$\Sigma_{(2,2)}=(1/\sqrt{2})\{\bar{\cal D}_+,{\cal D}_-\}$, in
terms of which the ${\cal N}=(0,2)$ chiral multiplet is given by
\be \Sigma=\left.\Sigma_{(2,2)}\right|_{\theta^-=\bar{\theta}_-=0}.
\ee
An ${\cal N}=(2,2)$ chiral multiplet $\Phi_{(2,2)}$ satisfies
$\bar{D}_+\Phi_{(2,2)}=\bar{D}_-\Phi_{(2,2)}=0$. This chiral
multiplet decomposes into an ${\cal N}=(0,2)$ chiral multiplet
$\Phi$ and a fermi multiplet $\Gamma$, defined by
\be \Phi &=& \left.\Phi_{(2,2)}\right|_{\theta^-=\bar{\theta}^-=0}
\nn\\
\Gamma&=&\frac{1}{\sqrt{2}}\,{\cal
D}_-\left.\Phi_{(2,2)}\right|_{\theta^-=\bar{\theta}^-=0}.\ee
If $\Phi_{(2,2)}$ transforms under a representation $R$ of the
gauge group, then both $\Phi$ and $\Gamma$ also transform under
$R$. A quick computation yields $\bar{\cal D}_+\Gamma=2i \Sigma
\Phi$, meaning that, in the notation of \eqn{e}, ${\cal N}=(2,2)$
supersymmetry imposes,
\be E=i\sqrt{2}\Sigma \Phi .\ee
The final ${\cal N}=(2,2)$ multiplet of interest is a twisted
chiral multiplet $\Sigma_{(2,2)}$, satisfying
$\bar{D}_+\Sigma_{(2,2)}=D_-\Sigma_{(2,2)}=0$. Like the ${\cal
N}=(2,2)$ chiral multiplet, this too decomposes into an ${\cal
N}=(0,2)$ chiral multiplet $\Sigma$ and a fermi multiplet $F$.
They are given by,
\be \Sigma&=&\left.\Sigma_{(2,2)}\right|_{\theta^-=\bar{\theta}^-=0}\nn\\
F&=&
-\frac{1}{\sqrt{2}}\bar{D}_-\left.\Sigma_{(2,2)}\right|_{\theta^-=\bar{\theta}^-=0} .
\ee
Note, however, that from the expansion \eqn{twistedc}, the
$\theta^+$ component of the ${\cal N}=(0,2)$ chiral multiplet
$\Sigma$ contains the barred fermion, rather than the unbarred
fermion,
\be \Sigma = \sigma - i\sqrt{2}\theta^+\bar{\zeta}_+
-i2\theta^+\bar{\theta}^+\partial_+\sigma .\label{02tc}\ee
This subtlety will prove important in what follows. Since twisted
chiral multiplets $\Sigma_{(2,2)}$ are always uncharged under the
gauge group, the corresponding fermi multiplet satisfies
$\bar{\cal D}_+F=0$.

\subsubsection{The Vortex Theory in ${\cal N}=(0,2)$ Language}

Let us finish this section by describing the ${\cal N}=(2,2)$
vortex theory of Section 2 in the language of ${\cal N}=(0,2)$
superfields. This will serve to fix notation for what is to come.
We decompose the fields as
\be \mbox{${\cal N}=(2,2)$ $U(k)$ Vector Multiplet}
&\longrightarrow& \mbox{$U(k)$ Vector Multiplet, $U$} \nn\\
&&+\ \mbox{Adjoint Chiral Multiplet
$\Sigma$}\nn\\
\mbox{${\cal N}=(2,2)$ Adjoint Chiral Multiplet} &\longrightarrow&
\mbox{Adjoint Chiral Multiplet, $Z$} \nn\\ &&+\
\mbox{Adjoint Fermi Multiplet $\Xi$} \nn\\
\mbox{${\cal N}=(2,2)$ Fund. Chiral Multiplets} &\longrightarrow&
\mbox{Fund. Chiral Multiplets, $\Phi_i$} \nn\\
&&+\ \mbox{Fund. Fermi Multiplets $\Gamma_i$}\nn\\
\mbox{${\cal N}=(2,2)$ Anti-Fund. Chiral Multiplet}
&\longrightarrow& \mbox{Anti-Fund. Chiral Multiplets, $\tilde{\Phi}_j$} \nn\\
&&+\ \mbox{Anti-Fund. Fermi Multiplet $\tilde{\Gamma}_j$}\nn\ee
where all the objects on the right are ${\cal N}=(0,2)$
superfields. As before, $i=1,\ldots, N_c$ for  $\Phi_i$ and
$j=1,\ldots N_f-N_c$ for $\tilde{\Phi}_j$. Appendix A contains a
list of the different component fields which appear in each of
these multiplets.

\para
The ${\cal N}=(2,2)$ supersymmetry imposes the relations,
\be \bar{\cal D}_+\Xi=2i[\Sigma,Z] \ \ \ ,\ \ \ \bar{\cal
D}_+\Gamma_i=2i(\Sigma-m_i)\Phi_i\ \ \ ,\ \ \ \bar{\cal
D}_+\tilde{\Gamma}^j=-2i(\Sigma-\tilde{m}_j)\tilde{\Phi}_j\ \ \ \
\ \label{22rels}\ee
(There is no sum over $i$ and $j$ on the right-hand side of these
equations). As we have seen, the right-hand side of each of these
equations appears as a potential ``$|E|^2$" arising in equation
\eqn{fermilag}. A further contribution to the worldsheet scalar
potential arises from the D-term, which provides the constraint
\eqn{vortexd}.

\section{The ${\cal N}=(0,2)$ Dynamics of Vortex Strings}

It is now time to present new results for the dynamics of vortex
strings in theories with ${\cal N}=1$ supersymmetry. Most of this
section is devoted to the  discussion of a simple deformation of
the ${\cal N}=2$ theory by the addition of a superpotential. In
Section \ref{more} we discuss a second class of  deformations.

\subsection{Adding a Superpotential}

We start by considering a ``Dijkgraaf-Vafa"-like deformation
\cite{dv}, breaking ${\cal N}=2$ to ${\cal N}=1$ through the
addition of a superpotential for the adjoint superfield $A$. The
superpotential now reads
\be {\cal W}=\sqrt{2}
\sum_{i=1}^{N_f}\tilde{Q}_i(A-m_i)Q_i+\hat{\cal
W}(A)\label{def}\ee
which gives rise to the scalar potential
\be V_{4d}&=& \frac{e^2}{2}\Tr(\,\sum_{i=1}^{N_f}\,Q_iQ_i^\dagger
- \tilde{Q}_i\tilde{Q}_i^\dagger - v^2{\bf 1}_{N_c})^2 +e^2\Tr|\,
\sum_{i=1}^{N_f}\tilde{Q}_iQ_i-\partial\hat{W}/\partial A|^2\label{v4dv}\\
&&+\sum_{i=1}^{N_f}\left(Q_i^\dagger
\{A-m_i,\bar{A}-\bar{m}_i\}Q_i +
\tilde{Q}_i\{A-m_i,\bar{A}-m_i\}\tilde{Q}_i^\dagger\right) +
\frac{1}{2e^2}\Tr|[A,A^\dagger]|^2 .\nn\ee
Let's look at how this superpotential affects the vacuum
structure. If ${\hat{\cal W}}$ is linear in $A$ then there is
merely  a constant piece in the F-term above and the Lagrangian
still preserves ${\cal N}=2$ supersymmetry. We can perform an
$SU(2)_R$ rotation of the scalar fields
$(Q_i,\tilde{Q}_i^\dagger)$ to bring the Lagrangian back to the
form \eqn{v4dn2}. We will assume that $\hW$ does not contain a
linear piece. In this case, for a generic superpotential $\hW(A)$,
$\tilde{Q}_i$ must turn on in the vacuum. Without loss of
generality, we choose the vacuum to be of the form,
\be Q^a_{\ i} = p_i\,\delta^a_{\ i} \ \ \ , \ \ \ \tilde{Q}^a_{\
i}=\tilde{p}_i\,\delta^a_{\ i}\ \ \ ,\ \ \ A={\rm
diag}(m_1,\ldots,m_{N_c})\label{wbac}\ee
with
\be  |p_i|^2-|\tilde{p}_i|^2=v^2 \ \ \ {\rm and} \ \ \
\tilde{p}_ip_i = \left.\frac{\partial\hW}{\partial a}\right|_{m_i}
\ \ \ \ \ \mbox{for each $i=1,\ldots,N_c$} \label{whatsleft}\ee

\subsection{What Becomes of the Vortex?}

Our goal is to understand how this deformation affects the
dynamics of the vortex string\footnote{Vortices in a  similar
system were studied in \cite{bolog}, but in the limit with
$v^2=0$, so that the vortex is built around a linear piece of
$\hW$. This gives rise to somewhat different physics from that
considered here.}. Let us firstly consider the case with distinct
masses $m_i$. Before adding the superpotential $\hW$ there were
$N_c$ different BPS vortices, each living in a different
$U(1)\subset U(N_c)$ and each with a different $Q_i$,
$i=1,\ldots,N_c$ carrying the asymptotic winding. What changes in
the presence of $\hW$?

\para
The crucial point to note is that something rather special happens
when the superpotential is tuned so that a critical point
coincides with one of the masses, say $m_k$ for some $k=1,\ldots,
N_c$
\be \left.\frac{\partial\hW(a)}{\partial
a}\right|_{a=m_k}=0 .\label{survive}\ee
If this is case, the vacuum equation \eqn{wbac} sets
$\tilde{Q}_k=0$. There is then no obstacle in constructing the
$k^{\rm th}$ vortex in which $Q_k$ winds; indeed the ${\cal N}=2$
vortex solution remains a solution in the deformed theory.

\para
Vortices of this type in ${\cal N}=1$ theories are often called
D-term vortices (the name arises because the symmetry breaking is
induced by a FI parameter, or D-term). It was shown in \cite{ddt}
that such objects are 1/2 BPS, preserving two of the four
supercharges of the four-dimensional ${\cal N}=1$ theory. In two
dimensions, there are two distinct superalgebras with two
supercharges: the non-chiral $(1,1)$ algebra, and the chiral
$(0,2)$ algebra. Given that the previous section was devoted to a
review of ${\cal N}=(0,2)$ theories, the reader may guess this
will be relevant for the vortex string. Let's now see that this is
indeed the case \cite{ddt}. The ${\cal N}=1$ supersymmetry
transformations for the vector multiplet fields are,
\be \delta A_\mu &=& -i\bar{\epsilon}\sigma_\mu\lambda +
i\bar{\lambda}\sigma_\mu \epsilon \nn\\ \delta
D&=&\bar{\epsilon}\bar{\sigma}^\mu {\cal D}_\mu \lambda + {\cal
D}_\mu \lambda \bar{\sigma}^\mu\epsilon \nn\\ \delta \lambda &=&
\ft12 \sigma^{\mu\nu}\epsilon F_{\mu\nu} + i\epsilon D .\ee
For each chiral multiplet $Q_i$, they take the form
\be \delta Q_i &=& \sqrt{2}\epsilon \psi_i\nn\\ \delta F_i&=&
i\sqrt{2}\bar{\epsilon}\bar{\Dslash}\psi_i -
2i\bar{\epsilon}\lambda Q_i\nn\\ \delta \psi_i &=&
\sqrt{2}\epsilon F_i + i\sqrt{2}(\!\Dslash Q_i) \bar{\epsilon} .\ee
Similar transformations also hold for the chiral multiplets
$\tilde{Q}_i$ with the appropriate substitutions. Finally, the
supersymmetry transformations for the adjoint chiral multiplet $A$
take the form,
\be \delta A &=& \sqrt{2}\epsilon \eta\nn\\ \delta F &=&
i\sqrt{2}\bar{\epsilon}\bar{\Dslash}\eta -
2i\bar{\epsilon}[\lambda, A]\nn\\ \delta \eta &=& \sqrt{2}\epsilon
F + i\sqrt{2}(\!\Dslash A) \bar{\epsilon} .\ee
The key point here is that the vortex equations \eqn{vort},
together with the requirement that $F_i=F=0$, provide solutions to
$\delta\lambda = \delta \psi_i = \delta\eta = 0$. The latter
condition $F=0$ is trivially satisfied when \eqn{survive} holds,
for then $\tilde{Q}_i=0$, while $A$ remains constant. To see which
supersymmetries are preserved in this case, it will suffice to
examine the $\delta \psi_i$ transformation. Using \eqn{ds}, in the
background of a stationary vortex so that ${\cal D}_+={\cal
D}_-=0$, we have
\be \delta\psi_{-i} = -2\sqrt{2}i ({\cal D}_z Q_i)
\bar{\epsilon}_- =0 \ \ \ {\rm and}\ \ \ \delta\psi_{+i} =
2\sqrt{2}i ({\cal D}_{\bar{z}}Q_i)\bar{\epsilon}_+ .\ee
In the background of a vortex, with the scalar field satisfying
${\cal D}_zQ_i=0$, we learn that $\bar{\epsilon}_-$ is the
preserved supersymmetry; it descends to provide the supersymmetry
variation parameter on the worldsheet. Meanwhile,
$\bar{\epsilon}_+$ is the broken supersymmetry which generates a
single  Goldstino mode on the worldsheet. In our notation
\eqn{goldy}, we have $\bar{\epsilon}_+=\chi_+/4$. (The $\chi_-$
collective coordinate in \eqn{goldy} arises from the second
supersymmetry transformation of the ${\cal N}=2$ theory. Its fate
in our ${\cal N}=1$ theory will be discussed shortly). The spinors
$\epsilon_\pm$ have definite, and opposite, chirality on the
worldsheet. This is the statement that the worldsheet theory
preserves chiral ${\cal N}=(0,2)$ supersymmetry, rather than
${\cal N}=(1,1)$.

\para
We have seen that, in the special case that a critical point of
$\hW$ coincides with a mass \eqn{survive}, there exists at least
one BPS vortex preserving ${\cal N}=(0,2)$ supersymmetry. But what
happens if this is not the case? If \eqn{survive} is not
satisfied, then there can be no BPS vortex solutions. To see this,
note that \eqn{whatsleft} tells us that $\tilde{Q}_k$  gains
an expectation value in the 4d vacuum. This means it cannot now
remain constant but, must wind asymptotically to ensure that its
kinetic term remains finite. A putative BPS vortex must now
satisfy,
\be {\cal D}_zQ_i={\cal D}_z \tilde{Q}_i=0 .\ee
Yet $Q_i$ and $\tilde{Q}_i$ have opposite charges.  A standard
theorem in mathematics
--- that a line bundle of negative degree has no non-zero
holomorphic section
--- states that there can only be simultaneous solutions to these
equations when either $\tilde{Q}_i=0$ or $Q_i=0$. (See, for
example, equation (3.43) of \cite{phases}). One can reach the same
conclusion by noting that $A$ is now also sourced in the vortex
background and $\delta\eta\neq 0$\footnote{The lack of BPS
vortices in this case is entirely analogous to the statement that
$F$-term vortices are not BPS in ${\cal N}=1$ theories \cite{ddt}.
In our set-up, the value of $\partial \hW/\partial a$ evaluated at
$a=m_k$ plays the role of the constant in the F-term in
\cite{ddt}.}. Of course, simple topological  arguments imply that
vortex strings still exist. However, they must satisfy the full
second order equations of motion, rather than the first order
Bogomolnyi equations, and their tension is strictly greater than
the BPS bound $T=2\pi v^2$.

\subsection{Vortex Dynamics}

In section 2, we described the ${\cal N}=(2,2)$ $U(k)$ theory on
the vortex worldsheet that captures the dynamics of $k$ parallel
vortex strings in  ${\cal N}=2$ four dimensional gauge theories.
We would like to understand how the worldsheet theory reacts to
the superpotential $\hW(A)$, breaking the four dimensional
supersymmetry from ${\cal N}=2$ to ${\cal N}=1$. We have seen
above that the vortices in the theory with superpotential $\hW(A)$
are classically BPS, preserving ${\cal N}=(0,2)$ supersymmetry,
when equation \eqn{survive} holds; otherwise there are no BPS
vortices. We would like to see this from the worldsheet.

\para
In fact, there is a unique deformation on the vortex worldsheet
that preserves ${\cal N}=(0,2)$ supersymmetry and reproduces the
expected vacuum structure described above. Recall from Section 3.2
that superpotentials in ${\cal N}=(0,2)$ theories are constructed
from fermi multiplets. The only such multiplet with a suitable
transformation under the $U(k)$ gauge symmetry is $\Xi$,
containing $\chi_-$ and the complex auxiliary field $G_Z$. The
worldsheet deformation is given by the ${\cal N}=(0,2)$
superpotential,
\be S_{{\cal W}} &\equiv& -\frac{1}{\sqrt{2}} \Tr_k\left.\int
d\theta^+\ \Xi\,J(\Sigma)\right|_{\bar{\theta}^+=0} - {\rm h.c.}
\nn\\ &=& -\frac{1}{\sqrt{2}}\Tr_k\,\left.\int d\theta^+\
\Xi\,\frac{\partial\hW(\Sigma)}{\partial\Sigma}\right|_{\bar{\theta}^+=0}-
{\rm h.c.}\label{booty}\ee
(up to some overall, unfixed, constant of proportionality). Note
that a superpotential of this form is a viable holomorphic term
since $\bar{\cal D}_+\Xi=2i[\Sigma, Z]\equiv i\sqrt{2}E_\Xi$ and
\be \Tr\ E_\Xi J=\sqrt{2}\Tr\ \left([\Sigma,
Z]\,\frac{\partial\hW(\Sigma)}{\partial\Sigma}\right)=0\ee
which satisfies the requirement \eqn{ej}. In principle there could
also be $\sigma$-dependent deformations of the kinetic terms for
$\Lambda_i$ and $\Xi$. As is common in supersymmetric field
theories, we will have less control over these ``D-term"
deformations, but will see that the superpotential \eqn{booty}
captures much of the important physics.

\para
The deformation \eqn{booty} has implications for both the bosonic
and fermionic zero modes of the vortex strings. We defer a
discussion of the fermions to the next subsection; we start here
by studying the bosonic zero modes. The extra bosonic term on the
vortex worldsheet arising from \eqn{booty} is a
potential\footnote{A note on dimensions: In 4d, $[\hW(A)]=3$,
which ensures that the scalar potential has the correct
dimensions: $[|\partial \hW/\partial A|^2]=4$. In 2d the auxiliary
field has dimension $[\sigma]=1$, so that $[\partial
\hW/\partial\sigma]=2$. The presence of the vortex tension, with
$[T]=2$, means that the worldsheet scalar potential \eqn{kimmoy}
has the correct scaling for the two dimensional worldsheet.}, %
\be V_{2d}= \Tr_k\left(T|G_Z|^2 +
G_Z\frac{\partial\hW(\sigma)}{\partial\sigma} + {\rm h.c.}\right)
=\frac{1}{T}\Tr_k\left|\frac{\partial\hW(\sigma)}{\partial\sigma}\right|^2 .
\label{kimmoy}\ee
We will now show that this gives the expected vacuum structure by
studying the $k=1$ vortex theory in some detail; the extension to
$k>1$ then follows.

\subsubsection{An Example: $k=1$ with $N_f=N_c$}

To illustrate the role of the superpotential \eqn{booty}, let's
look at the familiar $k=1$ theory of a single vortex in the case
with $N_f=N_c$ flavors. As we discussed in detail in Section 2,
when $\hW=0$ the internal moduli space is ${\bf CP}^{N_c-1}$ with
$\phi_i$ providing homogeneous coordinates. Once we turn on the
superpotential $\hW$, the bosonic part of the worldsheet theory is
given by
\be {\cal L}_{\rm bose}= T|\partial_mz|^2 + \sum_{i=1}^{N_c}\left( |{\cal
D}_m\phi_i|^2 - 2|\sigma-m_i|^2|\phi_i|^2\right) +
D(\sum_{i=1}^{N_c}|\phi_i|^2-r) - \frac{1}{T}\left|\frac{\partial
\hW}{\partial \sigma}\right|^2 + \frac{\theta}{2\pi}u_{01} .\nn\ee
In the presence of distinct, non-zero masses $m_i$, this
worldsheet theory has a supersymmetric ground state (i.e. with
vanishing vacuum energy) at
\be |\phi_j|^2=r\delta_{ij}\ \ \ ,\ \ \ \sigma=m_i\ee
only if $\hW(\sigma)$ has a critical point at $\sigma=m_i$
\be \left.\frac{\partial \hW(\sigma)}{\partial
\sigma}\right|_{\sigma=m_i}=0 .\label{again}\ee
This coincides with the expectations of the previous section: BPS
vortices only exist when \eqn{again} holds.

\para
When the masses do not coincide with the critical points, and
there are no BPS vortices, the potential $|\partial\hW/\partial
\sigma |^2/T$ determines the vacuum energy of the vortex string.
One could try to compare this to the excess tension of the non-BPS
vortex string, above the bound $T=2\pi v^2$, but this
unfortunately suffers from the previously mentioned ambiguity in
classical wavefunction renormalization for $\chi_-$ which also
affects the coefficient in front of $|G_Z|^2$.

\para
If the hypermultiplet masses vanish, $m_i=0$, then the story is a
little different. We may now set $\sigma=0$ in the vacuum (recall
that we assumed $\hW$ does not contain a linear piece, so
$\sigma=0$ is guaranteed to be a critical point). The full ${\bf
CP}^{N_c-1}$ bosonic moduli space is now restored. This is in
agreement with expectations from four dimensions, where we may
happily construct any vortex string, built around the vacuum with
$A=\tilde{Q}_i\equiv0$. We conclude that the superpotential
$\hW(A)$ does not affect the bosonic zero modes in this case, a
point made previously in \cite{syss}. However, the superpotential
does still affect the fermi zero modes. We now turn to a study of
 these.

\subsection{Fermions}

We will study the fermions in the case with vanishing
hypermultiplet masses $m_i=0$. Of all the Dirac equations in
\eqn{diracs}, only that for $\eta$ is modified by the
superpotential. It now reads
\be -\frac{i}{e^2}\Dbarslash\eta-\frac{i\sqrt{2}}{e^2}
[A,\bar{\lambda}]-\sqrt{2}\tilde{Q}^\dagger_i\bar{\psi}_i-
\sqrt{2}\bar{\tilde{\psi}}_i{Q}_i^{\dagger} -
\frac{\partial^2\hW(A)}{\partial A^2}\bar{\eta}=0 .\ee
In the background of the vortex, we may again set
$A=\tilde{Q}_i=0$. This means that all right-moving fermi zero
modes
--- those donated by $\lambda_+$ and $\bar{\psi}_{+i}$ ---
remain the same as in the ${\cal N}=(2,2)$ case, given by
solutions to
\be \sqrt{2}{\cal D}_z\lambda_+ &=& -e^2\,Q_i\bar{\psi}_{+i}\ , \nn\\
\sqrt{2}{\cal D}_{\bar{z}} \bar{\psi}_{+i}&=&-
Q_i^\dagger\lambda_+\label{same} .\ee
If the lowest order term in the superpotential $\hW$ is cubic or
higher, then the left-moving fermi zero modes are similarly
unaffected. However, if the superpotential $\hW(A)$ includes a
quadratic mass term
\be \hW(A)=\mu_2 A^2 + \ldots \ee
then the equations for the left moving fermi zero modes become
\be \sqrt{2}i{\cal D}_{\bar{z}} \eta_- &=& -e^2
\bar{\tilde{\psi}}_{-i} Q_i^\dagger - \sqrt{2}\mu_2\bar{\eta}_-\nn\\
\sqrt{2}i{\cal D}_z \bar{\tilde{\psi}}_{-i} &=&
\eta_-Q_i .\label{newbo}\ee
These equations are no longer related to the bosonic zero mode
equations \eqn{bogzero}: this is to be expected since, in breaking
to ${\cal N}=1$ supersymmetry, we have lost the half of
supersymmetry which ensured the correspondence between bosonic
zero modes and left-moving fermionic zero modes. Nevertheless, as
stressed in \cite{syss}, the Dirac equations \eqn{newbo} must
still admit the same number of solutions as the equations with
$\mu_2=0$. This follows from the fact that the zero modes are
chiral on the worldsheet, and cannot gain a mass through a
deformation. For a single $k=1$ vortex in the $U(2)$ gauge theory,
\eqn{newbo} was analyzed in \cite{syss}, both perturbatively in
$\mu_2\rho$, as well as in the large $\mu_2$ limit.

\para
To summarize, we learn that the deformation leaves the fermi zero
modes untouched unless $\mu_2\neq 0$, in which case it deforms the
profile of the left-moving fermi zero modes only. However, the
number of zero modes on the worldsheet remains the same. Let us
now compare this with the predictions from the proposed worldsheet
deformation \eqn{booty}.

\subsubsection*{Implications for Worldsheet Fermions}

In the presence of the superpotential $\hW$, the fermionic terms
in the $U(k)$ worldsheet theory read\footnote{As we mentioned
previously, the deformation from ${\cal N}=2$ to ${\cal N}=1$ may
also induce a finite wavefunction renormalization of the
left-moving fermion kinetic terms. We will not consider this
here.}
\be {\cal L}_{\rm fermi}&=&2iT\ \Tr_k\left(\bar{\chi}_-{\cal
D}_+\chi_-+\bar{\chi}_+{\cal D}_-\chi_+\right) +
2i\sum_{i=1}^{N_c}\left(\bar{\xi}_{-i}{\cal
D}_+\xi_{-i}+\bar{\xi}_{+i}{\cal D}_-\xi_{+i}\right) \nn\\
&&
-\sqrt{2}\,\Tr_k\left([\bar{\chi}_-,[\sigma,\chi_+]]-[\bar{\chi}_+,[\bar{\zeta}_-,
z]]-[\bar{\xi}_-,[\bar{\zeta}_+,z]]\right) + {\rm h.c.} \label{phew}\\
&& -\sqrt{2}\sum_{i=1}^{N_c}\left( \bar{\xi}_{-i}\sigma\xi_{+i}
-\bar{\xi}_{+i}\bar{\zeta}_-\phi_i +
\bar{\xi}_{-i}\bar{\zeta}_+\phi_i\right) +
\Tr_k\left(\chi_-\frac{\partial^2 \hW(\sigma)}{\partial
\sigma^2}\bar{\zeta}_+\right) + {\rm h.c.}\ . \nn\ee
The ${\cal N}=(0,2)$ superpotential is responsible for only the
final term. Integrating out the auxiliary fermions $\zeta_\pm$
again gives  constraints on the dynamical fermions,
\be \sum_i\phi_i\bar{\xi}_{+i}+[z,\bar{\chi}_+]=0\ \  \ {\rm and}\
\ \ \ \sum_i\phi_i\bar{\xi}_{-i}+[z,\bar{\chi}_-]=
\frac{\partial^2\hW(\sigma)}{\partial \sigma^2}\chi_- .\ee
We see that the right-moving fermions are unaffected by the
superpotential, in agreement with the Dirac equations \eqn{same}.
Similarly, if $\hW$ has no quadratic term, so $\mu_2=0$, then the
left-moving constraints are also left unchanged if we set
$\sigma=0$ (we shall see the role played by a non-zero $\sigma$
shortly). However, when $\mu_2\neq 0$, setting $\sigma=0$ still
leaves deformed constraints on the left-moving fermions. For
example, in the case of a single $k=1$ vortex, the constraints
read
\be
\sum_{i=1}^{N_c}{\phi}_i\bar{\xi}_{-i}=\mu_2{\chi}_- .\label{02cons}\ee
It's worth making a comment on this point. In the ${\cal N}=(0,2)$
theory, we have defined the left-moving worldsheet fermions such
that their kinetic terms are diagonal:
$\bar{\chi}_-\partial_+\chi_-+\bar{\xi}_{-i}{\cal D}_+\xi_{-i}$.
The constraint \eqn{02cons} holds in this basis. It is always
possible to redefine the fermions so that the constraint
\eqn{02cons} reverts to the original ${\cal N}=(2,2)$ constraint
\eqn{fermcons},
\be
\bar{\xi}'_{-i}=\bar{\xi}_{-i}-\frac{\mu_2}{r}\bar{\phi}_i\chi_- \
\ \ \Rightarrow \ \ \ \sum_{i=1}^{N_c}\phi_i\,\bar{\xi}'_{-i}=0 .\ee
This will then lead to a non-diagonal form for the fermion kinetic
terms.

\para
It was argued in \cite{syss} that, even in the presence of the
four-dimensional superpotential $\hW(A)=\mu_2A^2$, the worldsheet
theory of the vortex string still retains ${\cal N}=(2,2)$
supersymmetry. This argument was based on the survival of the
left-moving fermi zero modes, and the lack of a suitable ${\cal
N}=(0,2)$ deformation of the ${\bf CP}^{N_c-1}$ sigma-model. We
disagree with this conclusion. The vortex worldsheet theory is not
described by a ${\bf CP}^{N_c-1}$ sigma-model, but rather by a
${\bf C}\times {\bf CP}^{N_c-1}$ sigma-model and, as we have seen,
there is a suitable deformation of the latter in which $\chi_-$,
the left-moving fermion in ${\bf C}$, mixes with $\xi_{-i}$.
Moreover, this mixing is necessary to correctly capture the
bosonic properties of the vortex with arbitrary superpotential and
masses. As we explained above, to see this mixing between $\chi_-$
and $\xi_{-i}$ from an explicit analysis of the fermions would
require us to solve the fermi zero mode equations \eqn{newbo}, and
take their overlap to determine both the kinetic terms and the
constraint condition for the Grassmann collective coordinates of
the vortex.

\subsection{Symmetries and Other Aspects}

We now discuss various further aspects of the worldsheet theory,
starting with an analysis of the symmetries. We will show that the
worldsheet superpotential has the correct properties under
R-symmetry transformations to be induced by the superpotential
$\hW(A)$. The addition of the superpotential $\hW(A)$ breaks both
the $U(1)_R$ and the $U(1)_V$ symmetries in four dimensions. If
the superpotential takes the form,
\be \hW(A) = \sum_{n=2} \mu_n A^n .\label{mun}\ee
Treating the parameters $\mu_n$ as spurion fields, the symmetry is
restored if $\mu_n$ carries charge $(2-2n, 2)$ under $U(1)_R\times
U(1)_V$. Let us check that these charges descend to the worldsheet
theory. The deformation \eqn{booty} once again destroys both
$U(1)_R$ and $U(1)_V$ on the worldsheet, this time through the
presence of the worldsheet fermi interactions. The final term in
\eqn{phew} is\footnote{The presence of $\bar{\zeta}_+$ in this
expression, rather than $\zeta_+$, is crucial in this analysis. It
follows from the component expansion \eqn{02tc} and ultimately
from the fact $\Sigma$ arises from the decomposition of a $(2,2)$
twisted chiral multiplet as opposed to a $(2,2)$ chiral
multiplet.},
\be \sum_n n(n-1)\mu_n \ \Tr_k\,\left(\chi_-\sigma^{n-2}
\,\bar{\zeta_+}\right) .\label{yippee}\ee
Examining the table in Section \ref{symmetries}, we see that the
$U(1)_R\times U(1)_V$ worldsheet symmetry is again restored if
$\mu_n$ is assigned charges $(2-2n,2)$, in agreement with the
analysis in four dimensions.

\para
Note that the $U(1)_Z$ symmetry on the worldsheet, which arises
from rotational invariance in the $z=x^1+ix^2$ plane, is left
unbroken by the deformation \eqn{yippee} as, indeed, it must be.

\subsubsection*{Discrete Symmetries}

One can also check that the deformation on the worldsheet is consistent
with the discrete symmetries of the bulk theory\footnote{We thank M. Shifman and
A. Yung for stressing the importance of this.}. We start by considering the action
of parity, defined by
\be P: x^i \rightarrow -x^i\ \ \ \ i=1,2,3\ee
The original ${\cal N}=2$ theory can be written in terms of Dirac
spinors. For example, the adjoint Dirac spinor is
$\Psi = (\lambda,\bar{\eta})^T$.  Parity maps $P: \Psi \rightarrow \gamma^0\Psi$, or
\be P: \lambda \leftrightarrow \bar{\eta} \ \ \ {\rm and}\ \ \ \
P: \psi_i \leftrightarrow \bar{\tilde{\psi}}_i\label{fp}\ee
while for the complex adjoint scalar $P: A\rightarrow A^\star$. (The
imaginary part is really a pseudoscalar). Because the $z=x^1+ix^2
\rightarrow -z$ part of the parity transformation can be undone by
the rotation $U(1)_Z$ on the worldsheet, we may restrict attention to the simpler
parity transformation $P: x^3\rightarrow -x^3$, with $x^1$ and $x^2$ untouched.
This is the parity action under which the vortex string remains
invariant. It must therefore descend to the worldsheet. Indeed, as we
reviewed in Section 2, $(\lambda,\psi)$ donate right-moving zero modes
$\chi_+$ and $\xi_{+i}$,
while $(\bar{\eta}, \bar{{\tilde{\psi}}})$ donate left-moving zero
modes $\chi_-$ and $\xi_{- i}$. So the action of parity \eqn{fp} in the
bulk also exchanges left and right-movers on the worldsheet.

\para
So much for the ${\cal N}=2$ theory. What happens in the presence
of the ${\cal N}=1$ deformation? This pure parity symmetry \eqn{fp} is broken in
the 4d theory because the interactions of $\lambda$ and $\eta$ are
different. This is also seen in our ${\cal N}=(0,2)$ worldsheet
theory where the interactions of left and right movers differ.

\para
The 4d ${\cal N}=2$ theory is also invariant under $CP$.
Under charge conjugation, $C: B\rightarrow -B$ and the vortex is
mapped onto the anti-vortex. So this cannot be a symmetry of the
worldsheet. However, under the particular parity transformation
\be P': x^2 \rightarrow  -x^2 \label{p1}\ee
with $x^1$ and $x^3$ invariant, we also have $B_3\rightarrow
-B_3$. Moreover, the complex coordinate $z$ transverse to the
vortex string is mapped to $P': z\rightarrow z^\star$. This ensures
that the bosonic vortex solution is invariant under $CP'$.
For example, we have
\be {\cal D}_z Q_i\ \stackrel{C}{\longrightarrow}\ {\cal
D}_{z}Q^\dagger\ \stackrel{P'}{\longrightarrow}\  {\cal D}_{\bar
z}Q^\dagger \ee
so the Bogomolnyi equation ${\cal D}_z Q=0$ remains invariant
under $CP'$. When acting on the fermions, $CP'$ sends Weyl spinors to their
complex conjugates,
\be CP': \psi_i\rightarrow -i\sigma_2\bar{\psi}_i\ \ \ , \ \ \ \
CP': \lambda\rightarrow -i\sigma_2\bar{\lambda}\ \ \ ,\ \ \ {\rm etc.} \ee
This symmetry also descends to the worldsheet, where it acts as
complex conjugation, as can be checked explicitly from the zero
mode expressions of Section 2. We have,
\be CP': \phi_i \rightarrow \bar{\phi}_i \ \ \ {\rm and}\ \ \
CP': \xi_{\pm i} \rightarrow \bar{\xi}_{\pm i}\ \ {\rm etc} \ee
Note that, just as $CP'$ in the 4d theory didn't exchange
$\lambda$ and $\bar{\eta}$, so this symmetry on the worldsheet
doesn't send left-movers to right-movers. This can be
traced to the fact that the action $CP'$ under which the string
is invariant doesn't affect $x^3$.

\para
Unlike the pure parity transformation, the $CP'$ symmetry survives
the deformation to ${\cal N}=1$ supersymmetry.  More precisely, the
symmetry survives if the parameters in the superpotential ${\cal W}=\mu_n A^n$
are real. Alternatively we can think of these parameters as transforming
under $CP': \mu_n\rightarrow \mu_n^\star$. The same behavior is seen
in the worldsheet theory. Invariance of the final term in \eqn{phew} requires
that $CP':\mu_n\rightarrow \mu_n^\star$, in agreement with the 4d analysis.

\subsubsection*{The Four-Fermi Term}

So far we have neglected the role of $\sigma$ on the string
worldsheet. In the ${\cal N}=(2,2)$ case, we saw that $\sigma$
correctly takes into account the effect of the Yukawa couplings in
four-dimensions, resulting in a four-fermi term \eqn{224fermi} on
the worldsheet. It will play the same role here. The equation of
motion \eqn{yukyuk} for the adjoint field $A$ is now changed by
the superpotential $\hW(A)$. Even if the superpotential has
$\mu_2=0$, so the profiles of both left and right-moving fermionic
zero modes are the same as in the ${\cal N}=2$ theory, the
solutions to the full equations of motion, including Yukawa
sources for $A$, will necessarily differ. We would expect this to
feed back into the worldsheet dynamics. As in the ${\cal N}=(2,2)$
case, it is difficult to determine this explicitly, but thankfully
the lifting of the zero modes is once again dictated by the
symmetries of the problem.

\para
Let's start by examining the simplest case, with $\hW(A)=\mu_2
A^2$, so that the fermionic constraint equation is given by
\eqn{02cons}. Integrating out $\sigma$ on the worldsheet once
again gives rise to a four-fermi term
\be {\cal L}_{\rm 4-fermi}=
-\frac{|\bar{\xi}_{-i}\xi_{+i}|^2}{(r+2|\mu_2|^2/T)}\label{024fermi}\ee
which, up to an overall rescaling, looks the same as the ${\cal
N}=(2,2)$ four-fermi term \eqn{224fermi}. However this is
deceptive, for the constraints \eqn{02cons} ensure that
\eqn{024fermi} now includes a component of $\chi_-$. Previously,
as we discussed in Section \ref{symmetries}, $\chi_-$ was
prohibited from appearing in the four-fermi term since it was a
Goldstino mode in the ${\cal N}=(2,2)$ theory. It loses this
protection in the ${\cal N}=(0,2)$ theory.

\para
If the superpotential contains quadratic and higher order terms,
then integrating out $\sigma$ results not only in a four-fermi
term on the worldsheet, but also in a slew of higher order fermion
lifting terms. These terms are an interesting prediction of the
deformation \eqn{booty}.

\subsubsection*{A Comment on Anomalies}

In Section \eqn{add}, we saw that additional fundamental ${\cal
N}=2$ hypermultiplets in four dimensions contributed extra zero
modes to the vortex string which were captured in the gauged
linear sigma model by adding $(N_f-N_c)$ chiral multiplets in the
anti-fundamental representation of the $U(k)$ worldsheet gauge
group.

\para
There exists a trivial generalization in the ${\cal N}=1$ theories
in which we add only four-dimensional chiral multiplets, instead
of full hypermultiplets. For example, the addition of  a single
four dimensional chiral multiplet $Q$, transforming in the ${\bf
N}_c$ of $U(N_c)$, will contribute both bosonic and fermionic zero
modes to the vortex string. These live in an ${\cal N}=(0,2)$
chiral multiplet $\tilde{\Phi}$ of the worldsheet theory,
transforming in the $\bar{\bf k}$ of $U(k)$. In contrast, the
addition of $\tilde{Q}$, transforming in the $\bar{\bf N}_c$ of
$U(N_c)$, will contribute only fermi zero modes, living in an
${\cal N}=(0,2)$ fermi multiplet $\tilde{\Gamma}$ which transforms
in the $\bar{\bf k}$ of $U(k)$.

\para
While the above observation is trivial, there is an interesting
corollary in the quantum theory. The four-dimensional theory with
unequal numbers of fundamental and anti-fundamental chiral
multiplets is inconsistent at the quantum level, suffering a gauge
anomaly. This inconsistency descends to the vortex worldsheet,
which also suffers a $U(k)$ gauge anomaly unless the number of
chiral multiplets $\tilde{\Phi}$ is equal to the number of fermi
multiplets $\tilde{\Gamma}$. It would be interesting to study
vortices in chiral, anomaly free four-dimensional gauge theories,
to see if there is a corresponding delicate anomaly cancellation
on the vortex worldsheet.

\subsubsection*{The SQCD Limit}

To reach the ${\cal N}=1$ SQCD limit of the four-dimensional
theory, we send $\mu_2\rightarrow \infty$ to decouple the adjoint
chiral multiplet $A$. On the worldsheet, this has the effect of
decoupling the $U(k)$ adjoint chiral multiplet $\Sigma$. At the
same time, the constraint on the left-moving fermions \eqn{02cons}
becomes simply $\chi_-=0$, which effectively removes the fermi
multiplet $\Xi$. The right-moving fermions on the worldsheet are
still constrained to obey $\bar{\phi}\xi_{+i}=0$ (in the case
$N_f=N_c$) while the left-moving fermions $\xi_{-i}$ are
unconstrained. Nonetheless, the theory appears to be free of
worldsheet gauge anomalies.

\para
In this limit, the four-dimensional theory develops an enhanced, chiral flavor
symmetry $S[U(N_f)\times U(N_f)]$, rotating left and right movers independently.
(The ``S'' here is to remind us that the overall $U(1)_B$ is part of the gauge
group).
In the presence of the FI parameter, this is broken spontaneously and the
surviving symmetry in the vacuum is,
\be S[U(N_c)\times U(N_f-N_c)] \times U(N_f) \times U(1)_R\ee
Here the $U(1)_R$ is the anomaly-free R-symmetry. The same
symmetry enhancement is also seen on the vortex worldsheet theory
proposed above. There is once again a particular choice for the
anomaly free R-current.

\para
There is an issue with the normalizability of the fermi zero modes
in this limit. As $\mu_2\rightarrow 0$, the Dirac equation for the
left-moving fermi zero modes become ${\cal
D}_z\bar{\tilde{\psi}}_{-i}=0$ which has only non-normalizable
solutions. This could be mirrored on the worldsheet by infinite
kinetic terms for $\Gamma_i$, of the type that we neglected in the
discussion above. Alternatively, one could add a suitable
deformation to the 4d theory, such as the meson field considered
in \cite{gsy}, which once again renders these zero modes finite.

\subsection{A D-Brane Construction}

\EPSFIGURE{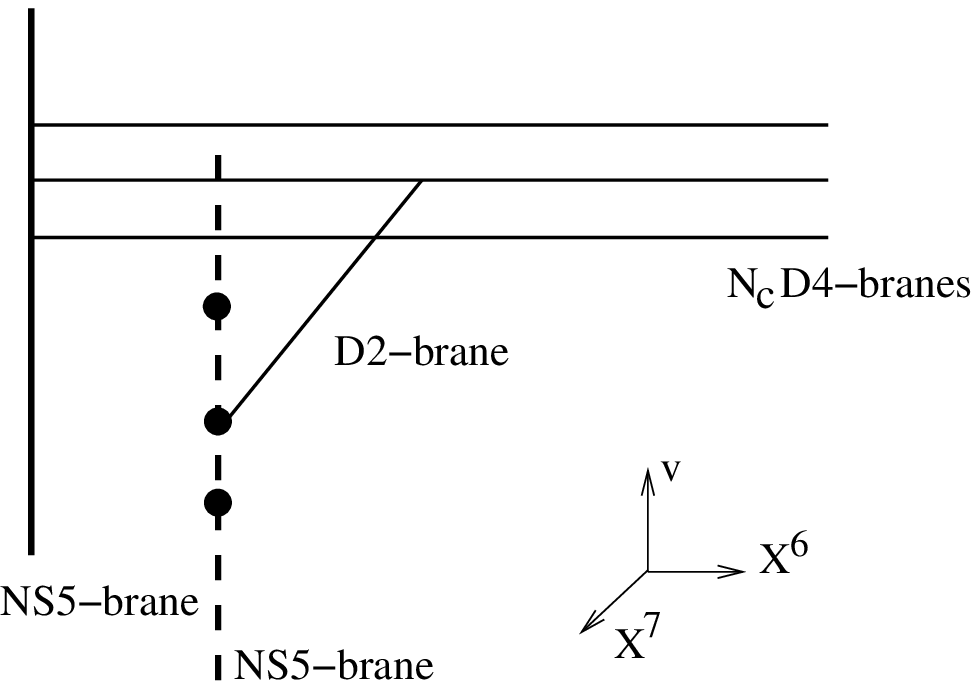,height=160pt}{}
One can construct a D-brane configuration whose low-energy
dynamics is governed by the four-dimensional theory of interest,
namely ${\cal N}=2$ super QCD, broken to ${\cal N}=1$ by the
addition of a superpotential $\hW(A)$ for the adjoint chiral
multiplet. One starts with the usual Hanany-Witten set-up for four
dimensional ${\cal N}=2$ gauge theories \cite{hanwit,w}. This
consists of two parallel NS5-branes lying in the $012345$
directions and separated a distance $\Delta X^6\sim l_s/e^2$ in
the $X^6$ direction. The ${\cal N}=2$ $U(N_c)$ gauge theory lives
on $N_c$ D4-branes, with worldvolume $01236$, which are suspended
between these two NS5-branes, while $N_f$ D6-branes with
worldvolume $0123789$ provide the hypermultiplets. To describe the
deformation \eqn{def} to ${\cal N}=1$ supersymmetry, we introduce
the complex coordinates
\be v= X^4+iX^5\ \ \ ,\ \ \ w=X^8+iX^9 .\ee
A superpotential $\hW(A)$ is induced on the D4-brane worldvolume
if we bend the right-hand NS5-brane so that it no longer lies at
the point $w=0$, but rather on the complex curve \cite{deboeroz}
\be w= \hW(v) .\nn\ee
Note that in the limit $\mu_n\rightarrow \infty$, with $\mu_n$
defined in \eqn{mun}, the curved NS5-brane becomes multiple flat
NS5-branes, lying a constant values of $v=X^4+iX^5$, given by the
roots of $\hW$. This is the description of the superpotential
first presented in \cite{first,second}.

\para
We may now pass through the series of moves described in
\cite{vib}, turning on a FI parameter by separating the two
NS5-branes in the $X^7$ direction, and identifying the vortices as
stretched D2-branes. The final result is shown in figure 3 in the
case of $N_f=N_c$. The figure shows a slice through $w=0$. The
dots depict the roots of $\hW(v)$, where the curved NS5-brane
intersects the $w=0$ plane; the ghostly dotted line shows where
the NS5-brane has left this plane and is living at some other
value of $w$. Figure 3 corresponds to a quartic superpotential,
with three critical points. One can check that the theory on the
D2-brane preserves ${\cal N}=(0,2)$ supersymmetry. It is clear
from the brane set-up that the D2-brane has a supersymmetric
ground state only when it may safely stretch from the curved
NS5-brane to a D4-brane, remaining at constant $v=m_i$ and without
leaving the safety of $w=0$. This requires
\be \left.\frac{\partial \hW(v)}{\partial v}\right|_{m_i}=0 .
\label{itsgood}\ee
This is the brane perspective on the statement that BPS vortices
only exist when \eqn{itsgood} is satisfied. It provides further
evidence that a worldsheet superpotential of the form \eqn{booty}
is required.

\subsection{A Different Superpotential}
\label{more}

To end this section, we consider a different deformation of the
${\cal N}=2$ theory which breaks the four dimensional
supersymmetry to ${\cal N}=1$. We add a superpotential of the
form,
\be {\cal W}_{{\cal N}=2} = \sqrt{2}\sum_{i=1}^{N_f}\,\tilde{Q}_i
{\cal V}_i(A) Q_i .\label{another}\ee
Here ${\cal V}_i(A)$ is an arbitrary holomorphic function of $A$.
The four-dimensional quantum dynamics of theories of this type was
previously studied in \cite{kap,seiji,rab,deboeroz}. We are here
interested in the effect on the vortex worldsheet. In fact, we
have already met one example of such a deformation that preserves
${\cal N}=2$ supersymmetry, because the complex mass term is of
this form with ${\cal V}_i(A)=A-m_i$. In that case, we saw that
the effect was not to induce a superpotential on the worldsheet,
but instead to change the relationship between $(0,2)$ fermi and
chiral fields,
\be \bar{\cal D}_+\Gamma_i=2i\Sigma\Phi_i\ \longrightarrow \
\bar{\cal D}_+\Gamma_i=2i(\Sigma-m_i)\Phi_i . \ee
Given this, it is natural to conjecture that the general
deformation \eqn{another} is captured by the worldsheet theory
with the relationship,
\be \bar{\cal D}_+\Gamma_i= 2i{\cal
V}_i(\Sigma)\Phi_i .\label{vsup}\ee
We will now provide evidence that this is indeed the case. We will
show that the deformation \eqn{vsup} is in agreement with all
symmetries of the theory, and reproduces the known behavior of the
vortex. The details of the calculations are similar to those
presented earlier, so we shall be brief.

\para
Let us firstly study what becomes of the vortex. We take the
vacuum of the four-dimensional theory to be
\be Q^a_{\ i}=v\delta^a_{\ i}\ \ \ ,\ \ \ \tilde{Q}_i=0\ \ \ \
A={\rm diag}(\nu_1,\ldots,\nu_{N_c})\label{4dvac}\ee
where $\nu_i$ is one of the roots of ${\cal V}_i$. If the $\nu_i$
are all distinct, the situation is the same as the one we
encountered in Section 2.5.1 with distinct masses $m_i$: there are
$N_c$ different vortices, each supported by the winding of a
different $Q_i$. In contrast, if all $\nu_i$ coincide, the full
${\bf CP}^{N_c-1}$ internal moduli space of the vortex is
restored.

\para
Let us see how this is reproduced on the vortex worldsheet by the
deformation \eqn{vsup}. For definiteness, we take a single $k=1$
vortex string in the $N_f=N_c$ theory. The bosonic part of the
worldsheet theory is given by,
\be  {\cal L}_{\rm bose}=T\,|\partial_m z|^2 +
\sum_{a=1}^{N_c}\left(|{\cal D}_m\phi^a|^2 - 2|{\cal
V}_i(\sigma)|^2|\phi_i|^2\right)
+D(\sum_{i=1}^{N_c}|\phi_i|^2-r)+\frac{\theta}{2\pi}u_{01} .\nn\ee
If the roots of $\nu_i$ of ${\cal V}_i(\sigma)$ are distinct, this
theory has isolated vacua, given by
\be |\phi_j|^2=r\delta_{ij}\ \ \ ,\ \ \ \sigma
=\nu_i .\label{tricky}\ee
However, there is an ambiguity here since ${\cal V}_i$ has
multiple roots $\nu_i$.  Suppose, for definiteness,  that ${\cal
V}_i(\sigma)$ is a polynomial of degree $P_i$. Then it appears
that, for each $i=1,\ldots, N_c$, there are $P_i$ different vacua
of the worldsheet theory. How are we to interpret these? In past
examples \cite{memono,sy,vstring}, different vacua of the
worldsheet corresponded to different physical vortices ---  see
Section 2.5.1. But we certainly don't want the same interpretation
here because the four-dimensional theory doesn't have $P_i$
distinct vortices, each with $Q_i$ winding asymptotically.
Thankfully, the interpretation of the multiple worldsheet vacua in
the present case is somewhat different. For fixed
$i=1,\ldots,N_c$, the $P_i$ different vacua differ only in the
value of the auxiliary field $\sigma$. The field $\sigma$ is to be
integrated out, set equal to its classical, algebraic equation of
motion. But there are $P_i$ different solutions to this algebraic
equation. The theory is only complete if we specify which of these
solutions we are to take. This means that the vacuum $\sigma =
\nu_i$ chosen in \eqn{tricky} is not a dynamical variable, but
rather a parameter of the worldsheet theory. We are therefore free
to fix it as we please, and the only natural candidate is to
equate it with the four-dimensional vacuum value $\nu_i$ in
\eqn{4dvac}\footnote{The equation of motion for $\sigma$ includes
a term  bilinear in the fermions, seen explicitly in \eqn{ria}.
The root of the equation of motion is taken to be the
four-dimensional vacuum value $\nu_i$ when the fermions vanish,
and is continuously connected to $\nu_i$ when the fermions turn
on.}. The end result is a situation where the same worldsheet
Lagrangian describes the vortex string in different
four-dimensional vacua; the specific four-dimensional vacuum of
interest appears as a boundary condition on the auxiliary $\sigma$
field.

\para
As a check of the conjecture \eqn{vsup}, we can confirm that the
$U(1)_R\times U(1)_V$ charges are consistent. If we write the
superpotential as
\be {\cal V}_i(A) = \sum_{n=0}\ h^{(i)}_n A^n\ee
then we are required to assign spurion charge $(2-2n,0)$ to
$h_n^{(i)}$. Let's check that this is in agreement with the
worldsheet. The deformation \eqn{vsup} gives rise to the terms
\be L_{\rm vortex} = \ldots + \sqrt{2}\sum_n \ (n h_n^{(i)}
\bar{\xi}_{-i}\sigma^{n-1}\phi_i\bar{\zeta_+}\ +\
h_n^{(i)}\bar{\xi}_{-i}\sigma^n\xi_{+i})+\ldots \label{ria}\ee
from which we learn that $h_n^{(i)}$ must again be assigned charge
$(2-2n,0)$ under the worldsheet $U(1)_R\times U(1)_V$.

\newpage
\section*{Appendix: The Alphabet}

This appendix is included to help the reader keep track of the
burgeoning conventions. The four dimensional fields are all
components of ${\cal N}=1$ superfields,
\be
\begin{array}{ll}
A_\mu: & \mbox{4d gauge field  in the vector multiplet $V$} \\
A: & \mbox{Adjoint valued 4d scalar in the chiral multiplet $A$} \\
Q_i: & \mbox{Fundamental 4d scalar  in the chiral multiplet $Q_i$} \\
\tilde{Q}_j: & \mbox{Fundamental 4d scalar in the  chiral
multiplet $\tilde{Q}_j$}
\\ \lambda: & \mbox{Adjoint valued 4d
fermion  in the vector multiplet $V$} \\
\eta: & \mbox{Adjoint valued 4d fermion in the chiral multiplet
$A$} \\ \psi_i: & \mbox{Fundamental 4d
fermion living in the chiral multiplet $Q_i$} \\
\tilde{\psi}_j: & \mbox{Anti-fundamental 4d fermion in the chiral
multiplet $\tilde{Q}_j$} .\end{array}\nn\ee
The worldsheet fields are all components of ${\cal N}=(0,2)$
superfields as described in Section 3:
\be
\begin{array}{ll}
z: & \mbox{Worldsheet scalar arising from broken translational
invariance,} \\ & \mbox{ in the chiral multiplet $Z$} \\
\phi_i: & \mbox{Worldsheet scalar corresponding to orientation
modes
of the string,} \\ & \mbox{in the chiral multiplet $\Phi_i$} \\
\sigma: & \mbox{Worldsheet auxiliary scalar in the  chiral multiplet $\Sigma$} \\
u_m: & \mbox{Worldsheet gauge field in the vector multiplet $U$} \\
{\chi}_+ & \mbox{Worldsheet Goldstino fermion  in the chiral
multiplet $Z$}
\\ \chi_- & \mbox{Worldsheet fermion in the fermion
multiplet $\Xi$} \\
\xi_{+i}: & \mbox{Worldsheet fermions living in the fermion
multiplet $\Phi_i$.} \\ \xi_{-i}: & \mbox{Worldsheet fermions  in
the fermion multiplet $\Gamma_i$} \\ \bar{\zeta}_+: &
\mbox{Worldsheet
auxiliary fermion living in the chiral multiplet $\Sigma$.} \\
\zeta_-: & \mbox{Worldsheet auxiliary fermion  in the vector
multiplet $U$} .

\nn\end{array}\ee

\section*{Acknowledgement}
We would like to thank Philip Argyres, Adam Ritz, and especially
Misha Shifman
and Alyosha Yung for helpful discussions. M.E. is supported in
part by DOE grant FG02-84ER-40153. D.T. is supported by the Royal
Society.

\end{document}